\def\gtorder{\mathrel{\raise.3ex\hbox{$>$}\mkern-14mu
    \lower0.6ex\hbox{$\sim$}}}
\def\ltorder{\mathrel{\raise.3ex\hbox{$<$}\mkern-14mu
    \lower0.6ex\hbox{$\sim$}}}

\documentclass[twocolumn]{aastex631}
\usepackage{graphicx}
\usepackage{blindtext}
\usepackage{amsmath}
\usepackage[normalem]{ulem}
\usepackage{soul}
\usepackage{natbib}
\begin{document}

\title{{Jetted Seyfert Galaxies at $z=0$:\\ Simulating Feedback Effects on Galactic Morphology and Beyond}}

\author[0009-0003-4860-8488]{Julianne Goddard}
\affiliation{Department of Physics and Astronomy, University of Kentucky, Lexington KY 40506-0055, USA}\thanks{E-mail: julianne.goddard@uky.edu}

\author[0000-0002-1233-445X]{Isaac Shlosman}
\affiliation{Department of Physics and Astronomy, University of Kentucky, Lexington KY 40506-0055, USA}
\affiliation{Theoretical Astrophysics, School of Sciences, Osaka University, Osaka 560-0043, Japan}
\thanks{E-mail: isaac.shlosman@uky.edu}

\author[0000-0002-0071-3217]{Emilio Romano-Diaz}
\affiliation{Argelander-Institut f\"ur Astronomie, Auf dem H\"ugel 71, 53121 Bonn, Germany}

\begin{abstract}
We use high-resolution cosmological zoom-in simulations to model feedback from Seyfert-type supermassive black hole (SMBH) jets onto galaxies with identical dark matter (DM) halos of log\,$M/M_\odot\sim 11.8$. The low mass, $\sim 10^6\,M_\odot$, seed SMBHs, have been introduced when the parent DM halos have reached log\,$M/M_\odot\sim 11$.  In a controlled experiment, we vary only the efficiency of the SMBH accretion and focus on galaxies and their immediate environment properties. Our results show that the AGN jet feedback has a substantial effect on the basic properties of Seyfert-type galaxies, such as morphology, gas fraction and distribution, star formation rate and distribution, bulge-to-disk ratio, DM halo baryon fraction, and properties of circumgalactic medium (CGM) and beyond. These have been compared to a galaxy with supernovae only feedback. We focus on the energy deposition by the jet in the ISM and IGM, and follow the expansion of the multiple jet cocoons to $\sim 2$\,Mpc. We find that the jet-ISM interaction gradually pushes the star formation to larger radii with increasing accretion efficiency, which results in increased mass of the outer stellar disk, which is best fit as a double-exponential disk.  Furthermore, we compare our galaxies and their properties with the observed nearby Seyfert galaxies, including the scaling relations, and find a close agreement, although statistical analysis of observed Seyferts is currently missing. In a forthcoming paper, we focus on evolution of these objects at $z\ltorder 10$ and study the effect of the SMBH seeding redshift.  
\end{abstract}



\section{Introduction}
\label{sec:intro}

Supermassive black holes (SMBHs) are ubiquitous in the universe and reside in galactic centers, serving as engines of active galactic nuclei (AGN), in the mass range of $\sim 10^6-10^{10}\,M_\odot$. About 10\% of AGN are radio loud and host relativistic jets \citep{begelman84}. The AGN generate energy across the electromagnetic spectrum, as well as mechanically, in the form of collimated (i.e., relativistic jets) and uncollimated outflows in the form of accretion disk winds. While jets are collimated by magnetic fields \citep[][]{blandford74,blandford82,contopoulos95}, the disk winds are either driven radiatively, e.g., by the UV line radiation \citep[][and refs. therein]{shlosman85,dekool95,murray95,arav97}, or magnetically \citep{blandford82,emmering92,konigl94,dekool95,bottorff97,elitzur06,fukumura14}, forming the broad emission and absorption line regions (BEL and BAL). Additional phenomena, like narrow line regions, warm absorbers, and powerful molecular outflows are the by-products of AGN activity which can deposit both energy and momentum in the host galaxy and beyond, in the galactic, circumgalactic (CGM) and intergalactic gas (IGM).

However, the essential details of the AGN feedback, e.g., how, where and in what way, the AGN energy is deposited, are still uncertain and subject to intense investigations. A significant observational evidence points to the AGN feedback curtailing the high-mass end function of galaxies, e.g., by quenching the star formation, while the stellar evolution feedback, e.g., supernovae (SN), determines the low mass-end function \citep{kauffmann03,mcnamara12}. 

In this work, we use high-resolution cosmological zoom-in simulations to investigate the mechanical feedback from relativistic jets in lower luminosity AGN, the Seyfert galaxies. We aim at understanding the effects of the energy and momentum deposition by these bipolar jets on their host galaxies, i.e., on the gas distribution in the ISM, CGM and IGM, star formation, stellar masses, the resulting galactic morphology, and other relevant parameters which characterize a galaxy. In the present work, we focus on a MW-type galaxy within a cosmological environment, at $z=0$. We seed the SMBHs and vary only the efficiency of the SMBH accretion, and thus the strength of the AGN feedback. These results have been compared to a model without the SMBH. Only SN feedback and AGN feedback from jets have been used here, in order to have a controlled experiment. In a subsequent work (Goddard et al., in prep., Paper\,II), we focus on the evolution of these galaxies at $z\ltorder 10$ and vary the seeding time.  A number of simulations have looked at the AGN jet feedback effects in Milky Way-type galaxies over cosmological timescales  \citep[e.g.,][]{byrne24,irodotou22,wellons23}, and we aim to expand this body of work.

Seyfert galaxies are classified as less luminous than quasars, i.e., with the bolometric luminosity of $L_{\rm bol}\ltorder 10^{12}\,L_\odot$ \citep[e.g.,][]{schmidt83} and SMBHs which are less massive, $M_\bullet\sim 10^6-10^8\,M_\odot$. Long-baseline arrays have resolved compact pc-scale jets in Seyferts and in low-ionization nuclear emission-line region (LINER) galaxies \citep[e.g.,][]{falcke00}.  While  short-baseline interferometry observations, e.g., VLA, have detected relativistic jets in Seyferts on the kpc-scales and larger \citep[e.g.,][]{ho01,rosario10}. Extended radio emission from Seyferts and LINERs has been explained by jet-ISM interaction \citep[e.g.,][]{gallimore06}.  High-velocity winds have been also  diagnosed in UV/optical \citep[e.g.,][]{komossa18}, similarly to quasars \citep[e.g.,][]{emmering92}. Half of the sample of 102 nearby radio-quiet Seyferts and LINERs show jetted structures, which is a higher fraction than in powerful quasars \citep[e.g.,][]{baldi18}. Bends and wiggles have been frequently observed in Seyfert jets connecting emission at different spatial scales, and jet precession has been invoked to explain them \cite[e.g.,][]{saikia09,hada13}.

A number of radio quiet Seyferts display jets on galaxy scale and larger, e.g., 2\,kpc jet in NGC4258  \citep{cecil00}, 12\,kpc jet in NGC 7479 \citep{laine08}, 5\,kpc in NGC4388 \citep{damas-segovia16}. While `normal' Seyferts in disk galaxies often host kpc-size jets \citep{keel06}, some stand out, like spectacular jets of 100\,kpc in J1649+263 \citep{mao15}, 350\,kpc in 0313-192 \citep{ledlow01}, 650\,kpc in Speca\,J1409-030 \citep{hota11}, 1.6\,Mpc in J2345-0449 \citep{bagchi14}, and more. Galaxies with large jets also display multiple episodes of launching them, e.g., NGC 2639, which exhibits relic radio lobes from 3\,pc to 9\,kpc \citep{rao23}, typically misaligned\footnote{These disk galaxies with such extended radio jets should be classified as radio galaxies rather than radio quiet Seyferts. Nevertheless, we emphasize that they are normal disk galaxies.}.

About 7\% of narrow line Seyfert\,1 (NLS1) galaxies are radio-loud and, therefore, harbor less powerful relativistic jets compared to radio quasars \citep[e.g.,][]{komossa06}. They accrete at a rate close to the Eddington \citep[e.g.,][and refs. therein]{komossa18}. But recent detections of jets in radio-quiet NLS1s have put into question the usage of radio loudness as an indicator of a relativistic jet whose radio emission could be shielded by the ionized gas \citep[e.g.,][]{berton20}.  The radio quiet NLS1s can be subject to low efficiency accretion \citep[][]{heckman14}.

To summarize, Seyfert host galaxies are typically radio quiet, but are jetted on all scales, from $\sim 1$\,pc to 100s of kpc, including multiple periods of jet activity. It is natural to ask the question, what effects jets in Seyferts have on the ISM of the host, the star formation, and ultimately on the morphology of these galaxies. Because accreting SMBHs have their mechanical and radiation output correlated with the SMBH masses \citep[e.g.,][]{kormendy95,magorrian98}, it is natural to expect that the most massive quasar-type SMBHs, $\gtorder 10^9\,M_\odot$, have the largest effect on their host galaxies. As Seyferts have lower luminosity than quasars and less massive SMBHs, they host weaker jets, which are oriented randomly with respect to the midplanes of galactic disks \citep{osterbrock82}, and so it is more difficult  to disentangle the influence of jets from stellar evolution in these objects.  

Jets interact with the environment by depositing both energy and momentum in the surrounding gas. For a collimated jet, its working surface, i.e., the `hot spot,' is relatively small, but the main interaction proceeds through its expanding cocoon. The cocoon consists of a shocked jet and ambient matter \citep{blandford74}. Its expansion along the jet, hence, results from the balance between the jet thrust and the ram pressure of the ambient matter. While in the perpendicular direction, the cocoon expands based on its internal pressure. Hence, the cocoon is overpressured with respect to the ambient gas and drives a shock \citep{begelman89}. Therefore, the cocoon can have a significant effect on the ISM, and, if it breaks out of the galaxy, then also on the circumgalactic and intergalactic medium, CGM and IGM, and on the associated processes there. 

In our modeling of the Seyfert jet interaction with the surroundings, we pay a special attention to the energy deposition by the jet, which has the most profound effect on the star formation. In low luminosity AGN, the feedback from the jets is more difficult to separate from the feedback from stellar evolution, such as stellar winds and SN.  

Efficiency of star formation in galaxies which form inside DM halos depends on many factors. Observations indicate that that stars form most efficiently per unit DM mass in the range of $M_{\rm h} \sim X\times 10^{12}\,M_\odot$. For lower or higher DM halo masses, this efficiency decreases \citep[e.g.,][]{behroozi19,piotrowska21}. Feedback from stellar winds and SN is responsible for this decrease in lower mass halos and correspondingly for lower mass galaxies embedded in these halos.  

For higher mass halos and associated more massive galaxies, the feedback from stellar evolution is probably insufficient to decrease the star formation efficiency. Instead, energy deposition by AGN is suspected to be responsible for this effect \citep{kauffmann03,mcnamara12}. From observations we do see that the prevalence of AGN peaks in a timeline consistent with the quenching of many massive galaxies \citep{kauffmann00}, which provides further evidence that these two phenomena are correlated.  Additionally, large-scale cosmological numerical simulations which include AGN feedback have been able to better match the observed stellar mass - halo mass relation in comparison to those without \citep[e.g.,][]{bourne23,rennehan24}. 

Higher resolution numerical simulations studying the influence of AGN feedback in massive galaxies and galaxy clusters have also shown that AGN are excellent candidates to slowdown and even halt the cooling flows, and quench the star formation on massive scales \citep{marti19,su21,komissarov21}.  Several of these studies have found that feedback in the form of jets is especially efficient at heating gas in the CGM and slowing the cooling flows, and that for galaxies with log($M_*$) $\geq$ 10 ${M_\odot}$ it may be the only way to fully quench the galaxy \citep{dutta24,scharre24}.

A number of scaling relations between AGN and their host galaxies have emerged based on observations, e.g., the $M_\bullet-\sigma$ relation, which relates the mass of the SMBH with the bulge velocity dispersion \citep{gebhardt00,ferrarese00}. Promisingly, numerical simulations have been able to replicate many of these observed scaling relations not only in massive galaxies \citep{dimatteo05}, but also in a broader range of galaxy masses. \citet{wellons23} looked at ($\sim 10^{10-13} M_{\odot}$) galaxies evolved over cosmological timescales with a wide range of AGN feedback mechanisms.  Of the many models they explored, collimated jet feedback was one of only a few that were able to match galaxy scaling relations, e.g., $M_*-M_{\rm halo}$, SFR-$M_*$, and $M_\bullet-\sigma$ relations, across the mass range. 

\citet{irodotou22} studied the effect of the 'radio-mode' feedback on MW-mass galaxies by modeling expanding bubbles in the CGM and compared this with isotropic mechanical feedback (e.g., winds), and with a no feedback model.  This study found that increasing feedback strength decreases galaxy stellar mass, and that varying the type of feedback can have a strong influence on the galactic morphology.  \citet{byrne24} used initially collimated jet feedback in MW-mass and intermediate mass galaxies along with other feedback prescriptions, and found that introducing feedback does reduce the star formation in both mass regimes. Additionally, they found that feedback is able to affect morphology, either increasing the half-mass radius of the galaxy or trending toward more spheroidal distributions with a stronger feedback. 

\citet{mukherjee18} and \citet{talbot22} simulated jets over relatively short timescales, $\ltorder 1$\,Gyr, in isolated Seyfert-type galaxies, and studied the jet-driven outflows and their effects on the galaxy and its environment.  \citet{talbot22} observed outflows launched into the CGM by low power jets and found that the jet direction  has a significant influence on the composition and velocity of the outflow. Jets that propagate into the disk of the galaxy tend to launch slower, colder outflows. \citet{mukherjee18} observed that effects on the SF depend on both the jet angle and its strength, as well as on the timeframe. Upon onset, the jets seem to trigger the SF, but over time they have dissolved cold clouds through enhanced turbulence, and the overall SF may start to decrease.  Additionally, stronger jets and those directed into the disk, are more likely to trigger the SF.
\citet{Appleby21} analyzed the effects of jet feedback on the CGM in the SIMBA simulations. They observed a reduced baryon fraction and increased metallicity of the hot gas in the CGM, due to the presence of jets.

There is a strong degeneracy in the AGN numerical recipes that are able to match observations, as follows from exploration of the parameter space of the AGN models \citep[e.g.,][]{su21,wellons23}.  A wide variety of AGN numerical recipes replicate observed galaxy scaling relations and efficacy in quenching the star formation.   

Due to the smaller sample of our models, we look in depth into the effects of the simulated jets on various parameters of evolving Seyfert-type galaxies over cosmological timescales, comparing them with the non-jetted one. We do verify the galaxy compliance with the major scaling relations. Our focus is on quantitative morphologies of modeled galaxies, on the bulge and disk components, distribution of star formation, accretion rates on the SMBH, effects on the baryonic components of galaxies, and the CGM. We follow the expansion of the cocoons generated by the jets and on the distribution of the energy deposition by the jets. In this paper we present the results at $z=0$, and in a forthcoming paper (Paper\,II), we shall present the cosmological evolution at $z\ltorder 10$.  The outline of this work is as following: section\,\ref{sec:numerics} deals with the numerical methods used, section\,\ref{sec:results} provides the results, which are discussed and summarized in section\,\ref{sec:discuss}.
 
\section{Numerics} 
\label{sec:numerics}

\subsection{Simulation setup}
\label{sec:sims}

We have performed a suite of cosmological zoom-in simulations of a single halo using the $N$-body/hydro code \textsc{gizmo} \citep{hopkins15}.  For better angular momentum conservation and in order to resolve the Kelvin-Helmholtz instability which is expected to develop when the cosmological filaments penetrate the DM halo, the MFM hydro solver was employed. The initial conditions (IC) of the simulations used the \citet{planck16} $\Lambda$CDM concordant model, with $\Omega_{\textrm m} = 0.308$, $\Omega_\Lambda = 0.692$, $\Omega_{\textrm b} = 0.048$, $\sigma_8 = 0.82$, and $n_{\textrm s} = 0.97$. We use the Hubble constant $h = 0.678$ in units of $100\,{\textrm {km}\,\textrm s^{-1}\,\textrm {Mpc}^{-1}}$. These were generated at redshift $z=99$ by using the \textsc{music} code \citep{hahn11} within a box of $50\,h^{-1}$\,Mpc, and were evolved to $z=0$.

From the parent, uni-grid, DM-only simulation, a set of halos was chosen for re-simulation at a higher resolution. DM halos and their properties were identified by the group finder \textsc{rockstar} \citep{behroozi12}, with a Friends-of-Friends (\textsc{FoF}) linking length of $b=0.28$. The DM halo virial radius and the virial mass, $R_{\textrm {vir}}$ and $M_{\textrm {vir}}$, have been defined by $R_{200}$ and $M_{200}$ \citep[e.g.,][]{navarro96}. $R_{200}$ is the radius within which the mean interior density is 200 times the critical density of the universe at that time, and $M_{200}$ the corresponding enclosed mass. For baryonic$+$DM halos we use $M_{\rm halo}$ and $R_{\rm halo}$ abbreviations. We selected a subset of halos with a final mass range $M_{\rm vir}$=$10^{11.5}-10^{12}\,M_\odot$, a local environment with relative overdensity of $\delta=2\pm 1.0$, and halo spin of $\lambda=0.03\pm 0.02$.  This paper focuses on the results using just one of the halos selected from this subset. The final properties of this halo are displayed in Table\,\ref{tab:DMsim}. Additional results involving other halos will be published elsewhere.

\begin{deluxetable}{ccccc}[ht!]
\tablecolumns{4}
\tablecaption{DM Halo Properties at $z=0$\label{tab:DMsim}}
\tablehead{
\colhead{Simulation Type} & \colhead{log\,$M_{\rm vir}/M_\odot$} & \colhead{$R_{\rm vir}$ [kpc]} & \colhead{$\lambda$} & $\delta$ }
\startdata {DM only} & 11.8 & 231 & 0.03 & 2.0 \\ 
{baryonic} & 11.8 & 230 & 0.02 & \\
\enddata
\tablecomments{The columns, from left to right, represent the DM halo virial mass, $M_{\rm vir}$, in the DM-only zoom-in and baryonic zoom-in simulations, the halo virial radius, $R_{\rm vir}$, in both simulation types, and the DM halo spin, $\lambda$.}
\end{deluxetable}

The zoom-in ICs are composed of five nested levels of refinement on top of the base grid, i.e., from $2^7$ to $2^{12}$. The DM-only version was first evolved in order to check for, and avoid contamination from massive, lower-resolution particles in the highest resolution-level volume. Subsequently, baryons were included at the highest level of refinement in the reconstruction of their respective ICs.  

Within this setup, the effective number of particles (DM and baryons) for our simulations is $2\times 4,096^3$, i.e., this is the number of particles under assumption that the entire simulation box has the resolution of the zoom region. This leads to a mass resolution per particle of $3.6\times 10^4\,{\rm M_\odot}$ for the gas and stars, and $1.9\times 10^5\,{\rm M_\odot}$ for the DM in the zoom region. The minimal adaptive gravitational softening in comoving coordinates for the gas is 1\,pc, for stars 20\,pc and 200\,pc for DM.

Galaxies have been identified by the group-finding algorithm \textsc{hop} \citep{eisenstein98}, using the outer boundary threshold of baryonic density of $10^{-4}\,n^{\textrm {SF}}_{\textrm {crit}} = 10^{-2}\,{\rm cm^{-3}}$, which ensured that both the host starforming gas and the lower density non-starforming gas are roughly bound to the galaxy \citep{romano-diaz14}. This assures that identified galaxies are not imposed with a particular geometry. 

\subsection{Star formation and SN feedback }
\label{sec:sf}
 
Gas heating and cooling from $10^{10}$\,K down to 10\,K are implemented, including H and He ionization$+$recombination, collisional, free-free, dust collisional, cosmic ray, and Compton effects, as well as metal-line \citep{wiersma09}, fine-structure, and molecular cooling, as detailed in \citet{hopkins18} and \citet{hopkins22}. 
Metal enrichment is included: the metallicity increases in the starforming gas and scales with the fraction of stars that turn into SN, and the metal yield per SN (see below). A total of 11 metal species were followed in both in gas and stars, including H, He, C, N, O, Ne, Mg, Si, S, Ca, and Fe. The H$_2$ abundances used for cooling calculations are estimated from the \citet{krumholz11} analytic fitting function.  

Metal diffusion is not implemented explicitly, but metals can be transported by mechanical feedback from SN and AGN (see below and section\,\ref{sec:smbh}). Our simulations include the redshift-dependent cosmic UV background \citep[e.g.,][]{faucher-giguere20,shen20}. 

The density threshold for star formation (SF) was set to $n^{\textrm {SF}}_{\textrm {crit}} = 100\,{\textrm {cm}^{-3}}$.  Stars form only where gas is self-gravitating, namely, when
\begin{equation}
        (\Delta v)^2+2c_{\rm s}^2 < 8 \pi G \rho\,.
\end{equation}
For this, we rely on the virial parameter, $\alpha_{\rm vir}$, defined as 
\begin{equation}
    \alpha_{\rm vir}=[(\Delta v)^2+2c_{\rm s}^2]/8 \pi G \rho\,,
\end{equation}
such that the self-gravity condition is met when $\alpha_{\rm vir} < 1$  \citet{hopkins13}. Star formation efficiency (SFE) is calculated from the virial parameter following the model by \citet{padoan12}:
\begin{equation}
    SFE= {\rm exp}(-1.4\alpha_{\rm vir}^{1/2}) ~.
\end{equation}
Once the SFE is determined, stars form stochastically \citep{springel03}.  Each star particle represents an entire population of stars with a mass distribution following the \citet{chabrier03} Initial Mass Function (IMF).

Mechanical feedback from SN--type\,II is implemented using the numerical prescription given in \citet{hopkins18}.  We follow individual SN events with an assumption that the SN occur with an IMF-averaged constant rate of $3\times 10^{-4}$ events per Myr and per $M_\odot$ for all stars less than 30\,Myr old.  Each SN event injects energy of $1\times 10^{51}$\,erg, mass of 14.8\,$M_\odot$, and the metal mass of 2.6\,$M_\odot$ into the surrounding gas within the radius of 200\,pc only.

\begin{deluxetable*}{cccccccc}[ht!]
\tabletypesize{}
\tablecolumns{8}
\tablecaption{Model Galaxies Properties at $z=0$\label{tab:galprops}}
\tablehead{
\colhead{Model} & \colhead{log\,$M_*/M_\odot$} & \colhead{log\,$M_{\rm gas}/M_\odot$} & \colhead{$f_{\rm gas}$} &\colhead{log SFR} & \colhead{log $Z_{\rm gas}$/$Z_\odot$} & \colhead{log $Z_*$/$Z_\odot$}\\
\colhead{$\epsilon$} &   &   & \colhead{} &\colhead{$M_{\odot}\,{\rm yr^{-1}}$} 
& \colhead{} & \colhead{} }
\startdata {$\epsilon_0$} & 10.97 & 9.95 & 0.09 & 0.30 & -0.77 & 0.04 \\ 
{$\epsilon_5$} &  10.93 & 9.82 & 0.07 & -0.44 & -0.98 & 0.05 \\
{$\epsilon_{15}$} & 10.66 & 9.98 & 0.17 & -0.77 & -0.81 & 0.00 \\
{$\epsilon_{50}$} & 10.58 & 9.52 & 0.08 & -2.72 & -0.84 & -0.02\\
\enddata
\tablecomments{Columns: (1) model name; (2) stellar mass inside HOP-defined galaxies; (3) gas mass inside HOP-defined galaxies; (4) gas fraction $f_{\rm gas}=M_{\rm gas}/(M_{\rm gas}+M_{*})$; (5) SFR averaged over the final 30\,Myr ($\epsilon_{50}$ is averaged over $60\,M_\odot$ to obtain a signal); (6) gas metallicity; (7) stellar metallicity.  }
\end{deluxetable*}

\subsection{The SMBH and its mechanical feedback }
\label{sec:smbh}

The SMBHs are modeled as sink particles that are seeded once the DM halo has reached $10^{11}\,M_\odot$.  They form from gas particles that are located at the minimum of the gravitational potential of the parent halo and inherit the initial mass, velocity, angular momentum, and position of the parent particle.  The SMBH also inherits angular momentum from the gas it accretes, so that the spin axis of the SMBH evolves with time.  We include an artificial velocity-damping term to continuously move the SMBH toward the most bound particle \citep[e.g.,][]{wellons23}. Without this, the SMBHs have a well documented tendency to wander from the center, being initially less than one order of magnitude more massive than the high resolution DM particles.
 
The SMBHs are seeded with $M_\bullet\simeq10^6\,M_\odot$ and grow by accreting surrounding gas, with an accretion rate determined based on gravitational torques \citep{shlosman89}, calculated using \citet{hopkins11} method,
\begin{equation}
\begin{aligned}
 \dot{M}_{\rm grav} =\, & \alpha f_{\rm d}^{3/2}\bigg(\frac{M_\bullet}        
    {10^8\,M_\odot}\bigg)^{1/6} \times   \\
&   \bigg(\frac{M_{\rm d}(<R_{0})}{10^9\,M_\odot}\bigg)\bigg(\frac{R_0}{\rm     100\,pc}\bigg)^{-3/2} \bigg(1+\frac{f_0}{f_{\rm gas}(<R_{0})}\bigg)^{-1} ,
\end{aligned}
\label{eq:grav}
\end{equation}
where $\alpha$ is a normalization factor motivated by differences in the SF criteria and set to a default value of $\alpha=5$.  $R_0$ is the kernel radius of the SMBH, and roughly corresponds to the SMBH radius of influence, limited to a minimum value of 80\,pc. This is the radius inside which the other parameters, e.g., $M_{\rm d}$($<$R$_{0}$) --- stellar disk mass, $f_{ \rm gas}$($<$R$_{0}$) --- gas fraction, and $f_{\rm d}$ --- disk mass fraction of the total mass, are evaluated. Finally,
\begin{equation}
    f_0=0.31f_{\rm d}^2 \bigg(\frac{M_{\rm d}(<R_{0})}{10^9\,M_\odot}\bigg)^{1/3}.
\end{equation}

We multiply the accretion rate by an efficiency parameter, $\epsilon$, introduced by \citet{angles-alcazar17} to make $\dot M_{\rm grav}$ more realistic and to capture the effects of unresolved processes affecting gas inflow that are not addressed in the existing subgrid recipe. 

In this work, $\epsilon$ is  the sole parameter varied between the AGN models. We set it to $\epsilon=0$, 4.5\%, 15\%, and 50\%, and denote these models as $\epsilon_0$, $\epsilon_5$, $\epsilon_{15}$ and $\epsilon_{50}$, respectively.  So the final accretion rate is
\begin{equation}
    \dot{M}_\bullet=\epsilon\dot{M}_{\rm grav}.    
\end{equation}
The efficiency parameter $\epsilon$ scales down the accretion rate onto the SMBH, $\dot M_{\rm grav}$, defined by Eq.\,\ref{eq:grav}. Smaller $\epsilon$ decreases the jet power, i.e., lowering the feedback strength. Note however that the jet power depends also on $\dot M_{\rm grav}$, which varies from model to model even with identical initial conditions. When we refer to "AGN strength," or to "feedback strength," we refer solely to the value of $\epsilon$.
\newpage

The SMBH accretion rate measured as a fraction of the Eddington accretion rate, $\dot M_{\rm Edd}$, is 
\begin{equation}
    f_{\rm Edd}=\dot M_\bullet/\dot M_{\rm Edd},
\end{equation}
where $\dot M_{\rm Edd}=L_{\rm Edd}/c^2$, and $L_{\rm Edd}$ is the Eddington limit for $M_\bullet$. 

To model the AGN jets, we use the hyper-refined particle spawning \citep{torrey20}, modified by \citet{su21}. Particles of mass $10^3\,M_\odot$ are spawned at a fixed fraction of the SMBH kernel radius with an initial velocity of $3\times 10^4\,{\rm km\,s^{-1}}$ and a temperature of $10^{10}$\,K.  {This gives an initial energy per particle budget of 
about 22\% emitted as thermal energy and 78\%  as kinetic energy.}

These particles are launched along the spin axis of the SMBH with a mass loading $\eta$=0.1, meaning 10\% of the accreted gas is returned in the form of jet particles.  The launching rate is thus $\dot{M}_{\rm jet}=\eta\dot{M}_\bullet$, giving a jet  {mechanical} energy injection rate of 
\begin{equation}
    L_{\rm jet}=1/2 \eta\dot M_\bullet v^2,
\end{equation}
where $v$ is the jet particle velocity. Velocities of the spawned particles are initially perfectly collimated, i.e., launched with a zero opening angle along the spin axis of the SMBH. We refer to this feedback as a 'jet' feedback. Once created, the jet particles interact hydrodynamically in the same way as any of the other gas particles. When they decelerate to at least one fourth of their initial velocity, and enter the kernel radius of another gas particle in a head-on trajectory, the jet particles re-merge and the mass-weighted properties of the two particles are averaged.  Note that here and throughout the paper, when we refer to the jet luminosity we reference only the mechanical luminosity.  As mentioned above, the thermal component of the jet energy is non-negligable, and contributes to significant heating of the ISM and CGM, particularly when the jet particles merge. This is discussed further in section \ref{sec:CGM}.

\begin{figure*}[ht!]
\includegraphics[width=1\linewidth]{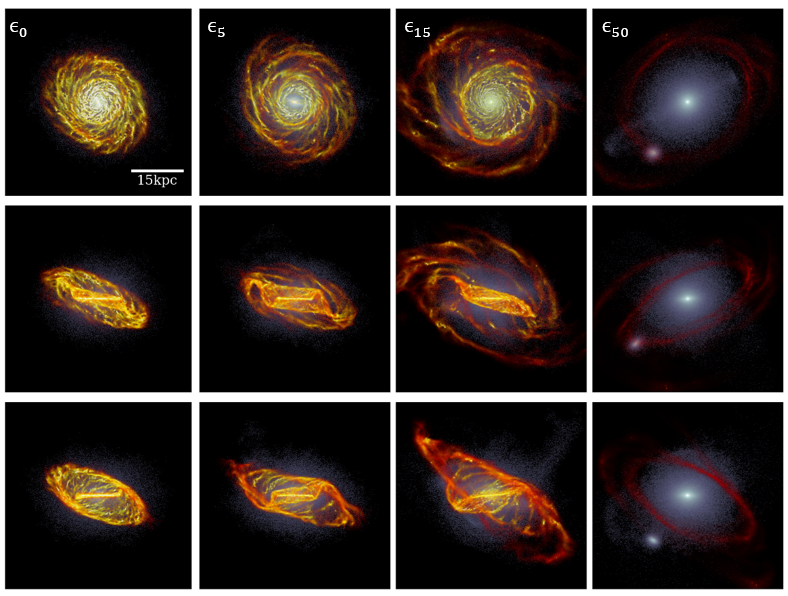}
\caption{Projected surface density of the HOP-selected galaxies at $z=0$ rotated based on the angular momentum, $J$, of the central 3\,kpc stellar disk. Orange color represents the gaseous component of a galaxy, while blue color represents the stars. The columns show the $\epsilon_0$, $\epsilon_5$, $\epsilon_{15}$, and $\epsilon_{50}$ models respectively from left to right. The top row images are the face-on projections, and the bottom two rows are the two perpendicular edge-on projections. Each frame is 54\,kpc $\times $ 54\,kpc.  
\label{fig:galaxies}}
\end{figure*}
    
\section{Results}
\label{sec:results}

Here we present the results based on our four model galaxies in their final state only, at $z=0$. All models have been evolved from the same initial conditions, and differ only in their AGN accretion efficiency, $\epsilon$. Paper\,II will focus on the cosmological evolution of the properties presented here. The SMBH has been seeded when the DM halo mass has reached $\sim 10^{11}\,M_\odot$ at $z=3.7$, in three out of the four models presented here. At this redshift, the stellar mass of the HOP-defined galaxy is $M_*\sim 1.3\times 10^{10}\,M_\odot$, and the gas mass of this galaxy is $M_{\rm gas}\sim 1.6\times 10^9\,M_\odot$, i.e., $\sim 12\%$. Prior to seeding, the SFR is $\sim 10\,M_\odot\,{\rm yr^{-1}}$.  

\subsection{Galaxies: global parameters at $z=0$}
\label{sec:scaling}

\begin{figure*}[ht!]
\includegraphics[width=1.0\linewidth]{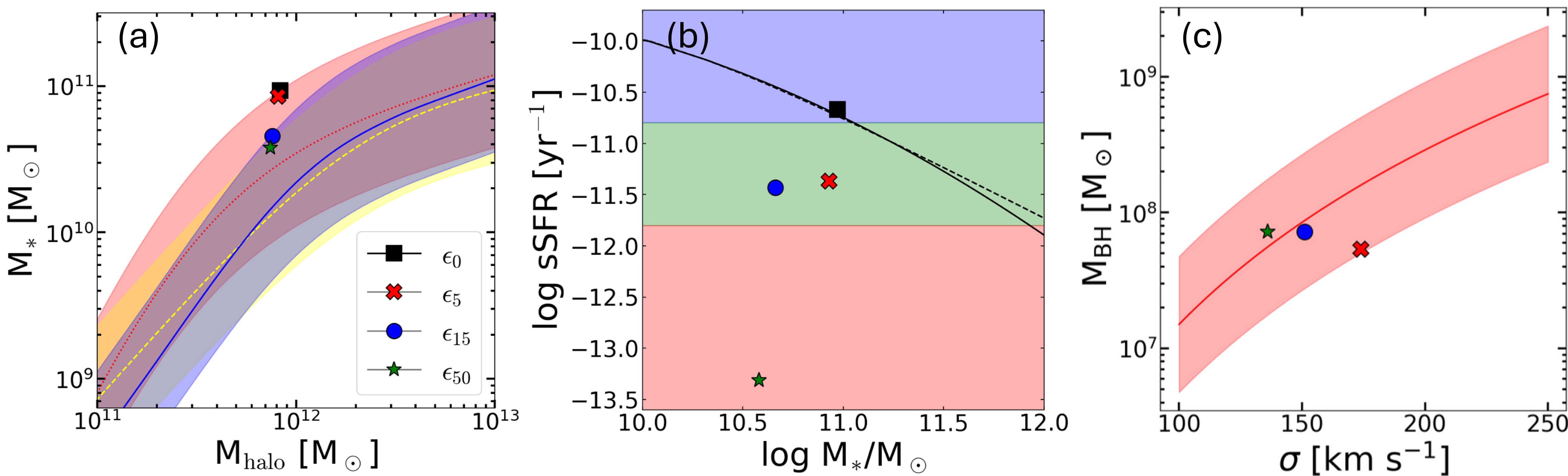}
\caption{Properties of our HOP-selected galaxies at $z=0$, as indicated in the legend. (a) Shows positions of modeled galaxies on the $M_*-M_{\rm halo}$ scaling relation adopted from \citet{behroozi19} (solid blue line), \citet{guo10} (dotted red line), and \citet{hudson14} (dashed yellow line). The filled regions around the medians correspond to a scatter of 0.5\,dex. (b) Provides specific SFRs for the HOP-galaxies and are plotted alongside the star-forming galaxy main sequence fits derived in \citet{Popesso23} (solid line --- their eq.10, dashed line --- their eq.14). The colors indicate divisions between the star-forming (blue), green valley (green), and quiescent (red), as delineated in \citet{salim14}. (c) The solid red line shows the $M_{\bullet}$ - $\sigma$ relation fit from \citet{kormendy13}, and the red fill shows the 0.5dex scatter. The 3-D dispersion velocity $\sigma$ has been corrected for the rotational velocity \citep{stewart22}.   
\label{fig:galprops}}
\end{figure*}

In Figure\,\ref{fig:galaxies}, we show the gas and stellar HOP-selected galaxies, on a scale $54\,{\rm kpc}\times 54$\,kpc, centered on the potential minimum of the galaxies. All of our models display a distorted gaseous component at this time. The source of this distortion and detailed analysis of the galaxy stellar components are discussed further in section \ref{sec:morph}. Only the $\epsilon_{50}$ model has an extended central cavity in the gas --- obviously the consequence of the AGN feedback, as we discuss here and in Paper\,II.  We also see evidence of recent gravitational interactions in the $\epsilon_{15}$ and $\epsilon_{50}$ models, both of these galaxies feature tidal stellar streams surrounding the galaxy, and the $\epsilon_{50}$ galaxy includes a small satellite that does not appear in the HOP-selected components of the other three models at this redshift.

The main properties of the simulated galaxies are listed in Table\,\ref{tab:galprops}. Those have been calculated within HOP-defined galaxies.  The total stellar mass within this radius is $\sim 10^{11}\,M_\odot$ for the $\epsilon_0$ model, and a factor of $\sim 2.5$ smaller for the highest AGN accretion, $\epsilon_{50}$, model. The stellar--mass decrease with increasing accretion efficiency is nearly monotonic. The gas fraction, defined as the gas mass in units of total baryonic galaxy mass, is $f_{\rm gas}\sim 0.1$. It shows a spread by a factor of 2, but in order to understand the dependency on $\epsilon$, one should follow the galaxy evolution over an extended time period. The gas distribution within the galaxy will be discussed later. 

The star formation rate (SFR), calculated over time intervals of 30\,Myr, exhibits a dramatic decline with $\epsilon$, by about three orders of magnitude. For $\epsilon_{50}$, this means essentially quenching the star formation process. The gas metallicity does not correlate with the accretion efficiency, and is smaller by a factor of a few compared to the stellar metallicity, which is about solar. This obviously points to being diluted by the gas accretion --- we return to this issue in section\,\ref{sec:ISM} and Paper\,II. 
 
Based on the global parameters of modeled galaxies, we have checked their positions on the stellar mass -- halo mass, $M_*-M_{\rm halo}$, diagram. Figure\,\ref{fig:galprops} (left frame) displays our models on this diagram adopted from \citet{behroozi19}, \citet{guo10} and \citet{hudson14}. All our models lie within 0.5\,dex above the median relation for the \citet{guo10} relation. A trend can be observed --- the models approach the median with increasing efficiency. Positions of the $\epsilon_0$ and $\epsilon_5$ models indicate that their feedback may not be strong enough to prevent over-cooling, and result in larger $M_*$. 

Over-cooling is a long-standing issue in numerical simulations \citep[e.g.,][and refs. therein]{shlosman13}, still in need of intense investigation, as it is found not only here, but in many other recent and similar bodies of work \citep[i.e.,][]{byrne24, chen23}. The addition of a stronger AGN accretion efficiency does decrease $M_*$ and brings it closer to the $M_*-M_{\rm halo}$ median, but understanding of the AGN feedback is far from satisfying.  

The $\epsilon_0$ model follows the SF main sequence, as indicated by the specific SFR (sSFR) provided in Table\,\ref{tab:galprops} and Figure\,\ref{fig:galprops} (middle frame). The colors in this frame indicate divisions between star-forming - blue, green valley - green, and quiescent - red, as delineated in \citet{salim14}. The final sSFR for $\epsilon_{15}$ and $\epsilon_5$ models fall within the green valley region, while the $\epsilon_{50}$ model is fully quenched. The values provided for the sSFR have been averaged over the last 30\,Myr of the simulation, except for the $\epsilon_{50}$ model, which has been averaged over 60\,Myr.  

The bulge masses and bulge velocity dispersions have been obtained from Sersic morphological decomposition (section \ref{sec:morph}), to determine the positions on the $M_{\rm bulge}-\sigma$ relation. The 3-D velocity dispersion has been corrected for contribution from rotational velocity \citep{stewart22}. The results for our galaxies are shown over-plotted on data from \citet{kormendy13} in Figure\,\ref{fig:galprops} (right frame). All AGN galaxies lie within the 0.5dex scatter region from the observational data. We elaborate on the evolution tracks of AGN galaxies in this diagram in Paper\,II. 

\subsection{Quantitative morphology}
\label{sec:morph}

The bulge and disk components of modeled galaxies have been determined by fitting the 1-D face-on stellar surface density profile of each galaxy within $0.1R_{\rm vir}$,  using the Sersic function \citep[e.g.,][]{bi22a} for the bulge and two exponential disks --- the double-exponential gave the best fits at $z=0$. We use the radii $0.1R_{\rm vir}$ in cylindrical shells instead of the HOP-defined galaxy size, as they are very similar and easier to calculate. The details of the fitting process are shown in appendix\,\ref{sec:appI} and in Figure\,\ref{fig:ssd} there. The summary of the fits are given in Table\,\ref{tab:deluxesplit}.

\begin{deluxetable*}{cccccccccccccc}[ht!]
\tablecolumns{14}
\tablecaption{Quantitative Morphological (Sersic) Decomposition of Modelled Galaxies at $z=0$ \label{tab:deluxesplit}}
\tablehead{
\colhead{Model} & n & \colhead{$R_{\rm b}$} & \colhead{$R_{\rm e}$} & \colhead{$R_{\rm disk1}$} & \colhead{$R_{\rm disk2}$} & \colhead{B/D} & \colhead{B/T} & \colhead{ log\,$M_{\rm bulge}/M_\odot$}& \colhead{log\,$M_{\rm disk1}/M_\odot$} & \colhead{log\,$M_{\rm disk2}/M_\odot$} & \colhead{log\,$M_{\rm tot}/M_\odot$} & $R_{\rm bar}$ & $e_{\rm max}$ \\
\colhead{$\epsilon$} & & kpc & kpc & kpc & kpc & &\colhead{} & & & & & kpc &
} 
\startdata {$\epsilon_0$} & 1.12 & 1.58 & 0.57 & 1.71 & 7.99 & 1.20 & 0.55 & 10.67 & 10.56 & 9.40 & 10.93 & 1.2 & 0.25 \\ 
{$\epsilon_5$} & 1.28 & 1.81 & 0.69 & 1.68 & 5.63 & 1.12 & 0.53 & 10.62 & 10.52 & 9.60 & 10.90 & 2.3 & 0.50 \\
{$\epsilon_{15}$} & 0.71 & 0.63 & 0.37 & 0.71 & 3.88 & 0.28 & 0.22 & 9.98 & 10.37 & 10.04 & 10.65 & 2.0 & 0.10 \\
{$\epsilon_{50}$} & 1.00 & 0.68 & 0.33 & 0.89 & 3.84 & 0.38 & 0.28 & 9.97 & 10.29 & 9.70 &  10.53 & -   & -  \\
\enddata
\tablecomments{Fitting the 1-D face-on stellar surface density profile using the Sersic function for the bulge and a double-exponential disk inside $0.1R_{\rm vir}$. Columns: (1) model name; (2) bulge Sersic index; (3) $R_{\rm b}$ bulge radius; (4) $R_{\rm e}$ bulge scalelength; (5) $R_{\rm disk1}$ inner disk scalelength; (6) $R_{\rm disk2}$ outer disk scalelength; (7) B/D bulge-to-total stellar disk mass ratio; (8) B/T bulge-to-total stellar mass ratio; (9) $M_{\rm bulge}$ bulge mass; (10) $M_{\rm disk1}$ inner stellar disk mass; (11) $M_{\rm disk2}$ outer stellar disk mass; (12) total stellar disk $+$ bulge mass; (13) $R_{\rm bar}$ bar or oval distortion radius; (14) $e_{\rm max}$ maximal ellipticity of a stellar bar or an oval distortion.}
\end{deluxetable*}

The bulge radii have been determined using the intersection of the fitting curves for each component. The $\epsilon_0$ and $\epsilon_5$ model bulges are similar, 1.58\,kpc and 1.81\,kpc. The models with a higher accretion efficiency, $\epsilon_{15}$ and $\epsilon_{50}$, have smaller bulges, 0.63\,kpc and 0.68\,kpc, respectively. The resulting bulge masses exhibit a monotonic decline, by a factor of 5, with increasing accretion efficiency. The inner stellar disk mass shows such a decline as well, albeit by a factor of 2 only. The outer stellar disk mass increases with $\epsilon$ by a factor of 5, except for the $\epsilon_{50}$ model, while the total stellar masses in galaxies decline by a factor of 3. Note that the total stellar mass of the HOP-selected galaxies corresponds closely to the stellar mass within $0.1R_{\rm vir}$. 

Table\,\ref{tab:deluxesplit} shows that lower accretion efficiency models, $\epsilon_0$ and $\epsilon_5$, have similar parameter values among themselves, with a dominant bulge, B/D$\sim 1.12-1.20$ (D being the stellar disk mass and B being the bulge mass.) The outer stellar disks are less massive than the inner ones by a factor of 10. The same trend, B/D$\sim 0.28-0.38$ is observed for the higher accretion efficiency, where the total disk dominates the system. The higher $\epsilon$ models possess smaller bulge components and disks that dominate in mass, and the outer stellar disks which are less massive than the inner ones only by a factor of less than 4.  

Hence, we conclude that a systematic difference in stellar morphology of modeled galaxies exists along the sequence of increasing accretion efficiency. Probably the most interesting trend is the appearance and increase in the mass fraction of the outer disk with increasing efficiency --- the SF is gradually pushed out to larger radii with the feedback.

\begin{figure}[ht!]
\includegraphics[width=1.0\linewidth]{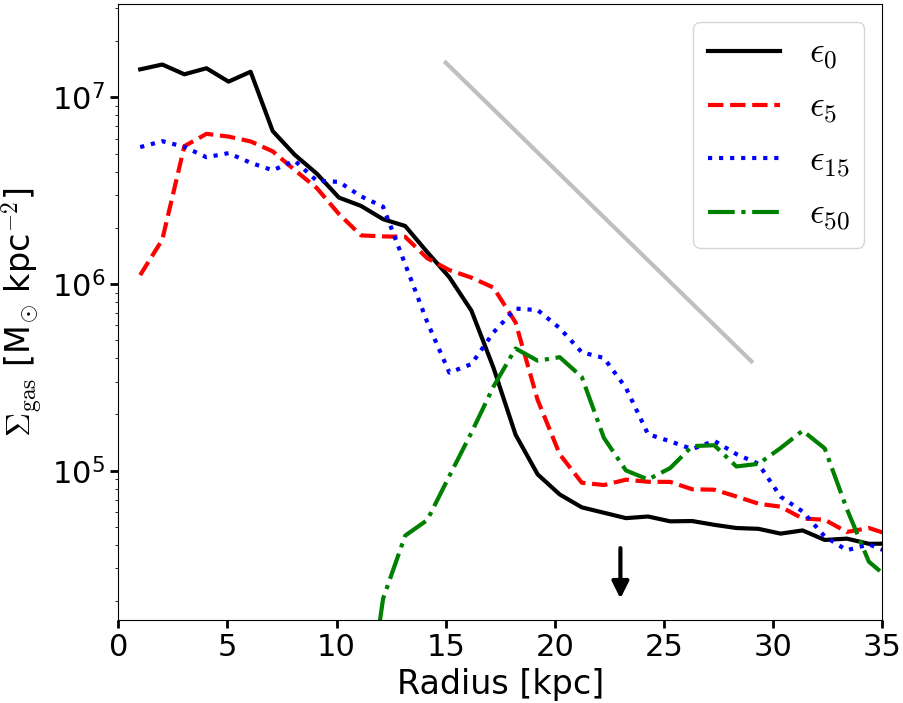}
\caption{Gas surface density of the face-on galaxy at z=0 measured in cylindrical shells of 1\,kpc width and $\pm 5$\,kpc height. The black arrow indicates $0.1R_{\rm vir}$. The offset gray line has been added to show an approximate exponential gas distribution with the scale length of 3.8\,kpc for comparison only.
\label{fig:profile}}
\end{figure}

We have also analyzed the presence of stellar bars. For this purpose, we have produced contour maps of stellar disk surface density and measured their ellipticity, $e(r)$. The bar size has been determined by identifying the radius of the maximal ellipticity, $e_{\rm max}$, and obtaining the radius of the subsequent 15\% decrease from this maximum \citep{marti06}.  Table\,\ref{tab:deluxesplit} shows $e_{\rm max}$ and $R_{\rm bar}$ for all models, if a bar or an oval distortion are present. Here we adopt the definition of a bar having $e_{\rm max}\ge 0.4$, which approximately corresponds to the ratio of a Fourier amplitude for $m=2$ mode, $A_2$, normalized by the monopole $m=0$ mode, $A_0$, and which is $A_2/A_0 = 0.15$ \citep{marti06,bi22b}. 

The $\epsilon_0$ model hosts an oval distortion rather than a stellar bar, with a maximal ellipticity of $e_{\rm max}\sim 0.25$ with a radius of 1.2\,kpc.   For comparison, $A_2/A_0\sim 0.1$, which justifies our conclusion of an oval distortion. Even a weaker oval distortion is present in the $\epsilon_{15}$ model, with $e_{\rm max}\sim 0.1$ of a radius of 2\,kpc. Not even an oval distortion is present in $\epsilon_{50}$ model. But the $\epsilon_5$ model displays a bar with $e_{\rm max}\sim 0.5$ with $R_{\rm bar}\sim 2.3$\,kpc. 

While only $\epsilon_5$ model exhibits a weak bar at $z=0$, other models display such bars at earlier times (Paper\,II). These bars and oval distortions are typically triggered by prograde tidal interactions, but the retrograde interactions can also weaken the bars. Additional factors, such as a rapid influx of gas along the cosmic filaments can have the same effect \citep{bi22b}.

\begin{figure*}[ht!]
\includegraphics[width=1.0\linewidth]{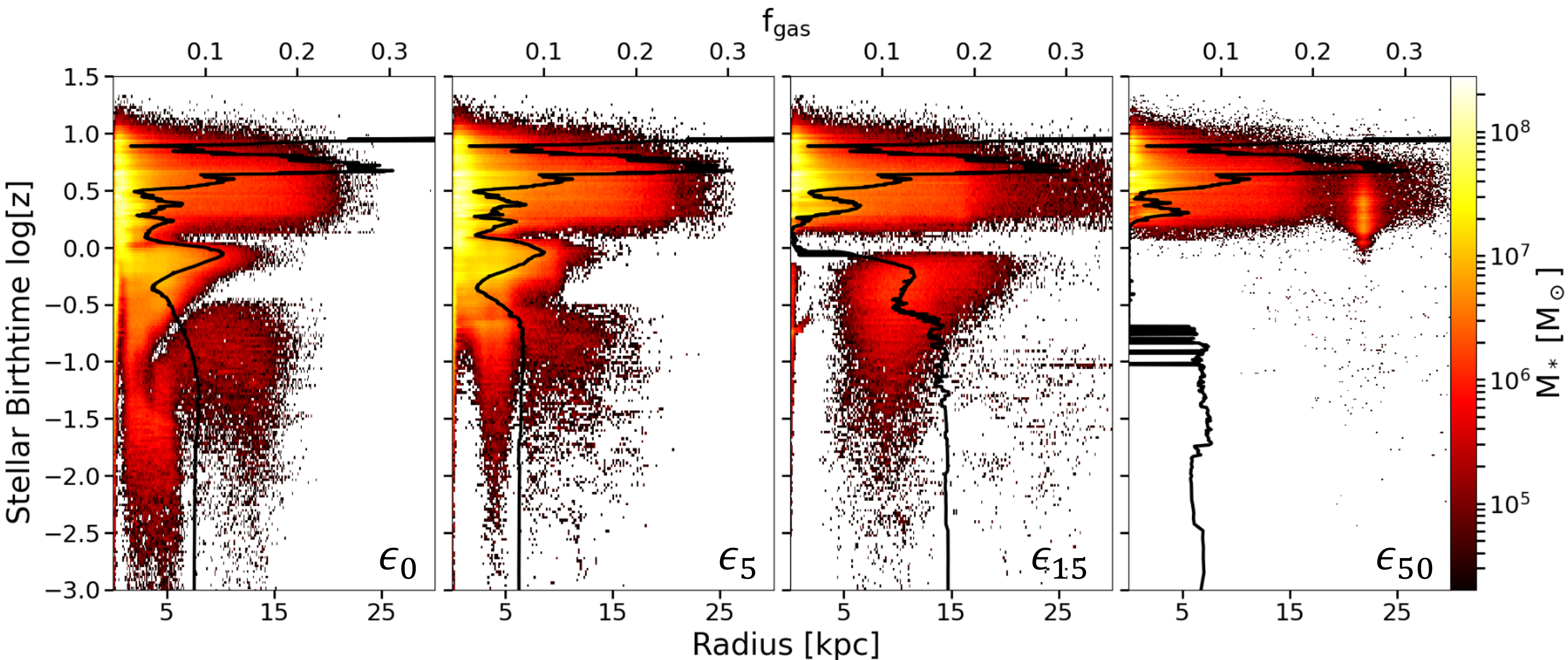}
\caption{2-D histogram of stellar age vs radius divided into $300\times 300$ grid cells with the color representing the stellar mass in each cell at $z=0$.  The overplotted solid black line represents the evolution of the gas fraction in galaxies.  The stars plotted here are those residing in the  HOP-galaxies at $z=0$, and have not necessarily formed within the galaxies. 
\label{fig:agevsrad}}
\end{figure*}

The radial 2D distribution of gas in the models, $\Sigma_{\rm gas}$, is more diverse than the stellar distribution, and is shown in Figure\,\ref{fig:profile}. Among all the models, the $\epsilon_0$ has the highest surface density in the central few kpc, and the lowest one in the outer galaxy. In the central $\sim 5$\,kpc, the gas surface density forms a plateau, with $\epsilon_0$ being the highest. The $\epsilon_5$ model shows a sharp decline in the central 3--4\,kpc which is the result of the combined action of the stellar bar and the AGN feedback. Bars funnel the gas inwards, contributing to the gas deficiency within their radii \citep[e.g.,][]{shlosman89,heller94}. The $\epsilon_{50}$ model displays the strongest departure from the stellar surface density, with most of the gas being pushed to the galaxy edge, as seen in Figure\,\ref{fig:galaxies}. This effect is also observed in $\epsilon_5$ and $\epsilon_{15}$ models, whose gas surface density outside $\sim 15$\,kpc exceeds that of $\epsilon_0$. It affects also the distribution of the SF in the disk, and we return to this point in the next section.

\subsection{Star formation}
\label{sec:starForm}

We can further examine the distribution of stars in Figure\,\ref{fig:agevsrad}. For all the models, about half of the stars have been formed by $z\sim 2$, and another third by $z\sim 1$. This older population of stars seems to not only populate the bulges, but also the disks, as we observe them at all radii. The low AGN efficiency models have stars residing out to $\sim 20-25$\,kpc, while for high AGN efficiency models, some stars reach to almost 30\,kpc. These stars which are born at large radii belong to the stellar halo population and have been accreted during cosmological evolution. The amount of younger stars decreases with increasing effciency, confirming that the $\epsilon_{50}$ model has a quenched SF. 

\begin{figure}[ht!]
\includegraphics[width=0.95\linewidth]{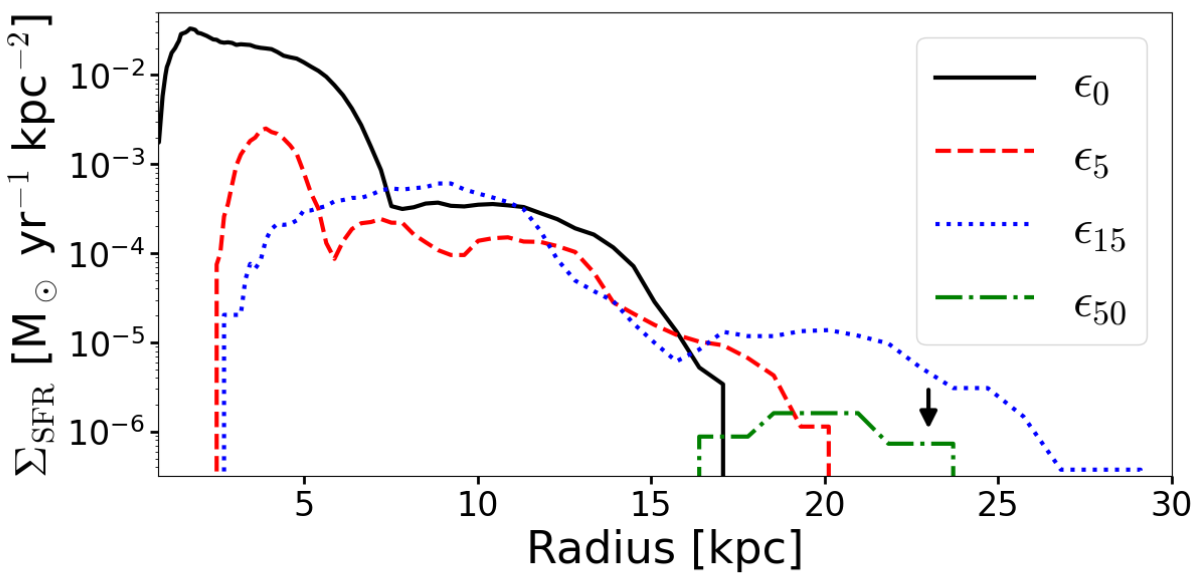}
\caption{The surface density of the SFR vs radius averaged in cylindrical shells of height $\pm 5$\,kpc and 20 logarithmically spaced bins in radius out to 30\,kpc. $\Sigma_{\rm SFR}$ is averaged over the last 100\,Myr of the simulation to increase the signal and minimize the effects of transient behavior in the star formation. The black arrow indicates the radius corresponding to 0.1R$_{\rm vir}$.
\label{fig:SFRSSD}}
\end{figure}

\begin{figure*}[ht!]
\includegraphics[width=0.95\linewidth]{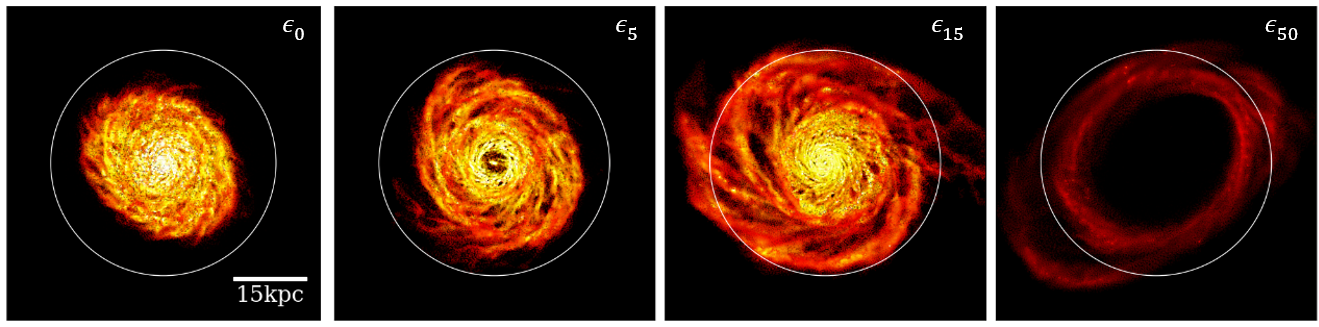}
\caption{Face-on gas morphology of the HOP-selected galaxies at $z=0$. The white circles  display the $0.1R_{\rm vir}$ radius for comparison. The central cavity for the $\epsilon_5$ model is the gas response to the stellar bar. The gas in the $\epsilon_{50}$ model has been pushed away from the stellar disk, which explains the quenching of SF there. Note also the gradual increase in the gaseous disk size with increasing efficiency. 
\label{fig:gasDisk}}
\end{figure*}

Figure\,\ref{fig:agevsrad} shows that the distribution of stellar ages correlates strongly with the gas fraction in the host galaxy. This is most visible at $z\sim 1$, where we observe a gap in the stellar age population, which coincides with the minimum in the gas fraction at that time. The prominent gap in stellar age distribution is visible in all models at $z\sim 1$. The gap becomes more prominent with AGN efficiency, and for $\epsilon_{50}$, basically determines the quenching of the SF. This reduction in SF seems to be triggered by a series of minor interactions, which substantially reduce the gas fraction in galaxies, and is subsequently enhanced by the stellar and AGN feedback. While gas fraction and thus the SFR are reduced in all models, for the stronger efficiency AGN models the feedback works in tandem to cause a more severe and sometimes a longer-lived gas loss. We also observe that the stars formed after $z\sim 1$ gradually avoid the central few kpc with increasing efficiency. They have formed in the outer disk or been brought in by minor and intermediate mergers (assuming that major mergers would destroy the stellar disk.)   

Interestingly, the gas fraction in galaxies correlates with the stellar birth time, only with a large time delay, as seen in Figure\,\ref{fig:agevsrad}. While the gas fraction declined from $f_{\rm gas}\sim 0.25$ to $\ltorder 0.05$ around $z\sim 3$, because of numerous interactions and minor mergers, the SFR declined only around $z\sim 1.5$.

Note also that the gas fraction, $f_{\rm gas}$, varies by a factor of a few over $\sim 3$\,Gyr time periods after $z\sim 1$, strongly correlating with the stellar age distribution. Finally, $f_{\rm gas}$ stabilizes during the last 2\,Gyr.   

Figure \ref{fig:SFRSSD} provides the radial distribution of the SFR, averaged over the last 100\,Myr, and expressed as the surface density of the SFR. We observe that while the $\epsilon_0$ model SFR peaks close to the center, at $\sim 1$\,kpc, models with higher $\epsilon$ peak progressively at larger radii. The $\epsilon_5$ model peaks at about 5\,kpc, and $\epsilon_{15}$ model at $\sim 10$\,kpc. The $\epsilon_{50}$ model shows a very low signal and only between 20--25\,kpc, at the galaxy outskirts. This distribution of the SFR is well correlated with Figure\,\ref{fig:agevsrad}. 

The gas distribution in modeled galaxies can also be visualized in Figure\,\ref{fig:gasDisk}. It displays an $\epsilon$-sequence, i.e., based on the AGN accretion efficiency strength. The most striking observation is the dependency of the gaseous disk size and its activity of spiral arms on this efficiency. The $\epsilon_0$ model has a gaseous disk located well inside the $0.1R_{\rm vir}$ radius. The $\epsilon_5$ model exhibits a larger disk, supplemented with increased activity of the outer spirals. The $\epsilon_{15}$ model has a disk filling up the $0.1R_{\rm vir}$ radius and extensive spiral arms outside this radius. These models are in sharp contrast with the $\epsilon_{50}$ model. Here, the gaseous disk is absent and replaced by a gaseous ring at $0.1R_{\rm vir}$. 
 
\begin{deluxetable}{cccccc}[ht!]
\tabletypesize{}
\tablecolumns{6}
\tablecaption{The SMBH-Related Properties at $z=0$} \label{tab:BHprops}
\tablehead{
\colhead{Model} & \colhead{log\,$M_\bullet$} & \colhead{log\,$\dot{M}_\bullet$} & \colhead{log\,$f_{\rm Edd}$} & \colhead{log\,$L_{\rm jet}$}  & \colhead{Jet Angle} \\
\colhead{$\epsilon$} & \colhead{$M_\odot$} & \colhead{$M_\odot\,{\rm yr^{-1}}$} &  & \colhead{${\rm erg\,s^{-1}}$} &  \colhead{degrees}
} 
\startdata {$\epsilon_5$} & 7.73 & -2.20 & -2.28 & 41.25 & 20.7 \\
{$\epsilon_{15}$}         & 7.86 & -2.59 & -2.79 & 40.86 & 37.0 \\
{$\epsilon_{50}$}         & 7.86 & -3.26 & -3.47 & 40.19 & 51.5 \\
\enddata
\tablecomments{Columns: (1) model name; (2) $M_\bullet$ the SMBH final masses; (3) $\dot M_\bullet$ the SMBH accretion rates; (4) the SMBH accretion rates as a fraction of the Eddington rate; (5) $L_{\rm jet}$ the jet mechanical luminosity; (6) the jet angle with the disk spin axis. The galaxy spin axis is defined by the stellar angular momentum vector within the inner 3\,kpc.  The angle, mechanical luminosity, and accretion rate have been averaged over the final 30\,Myr of simulations. }
\end{deluxetable}

\begin{figure*}[ht!]
\includegraphics[width=1.\linewidth]{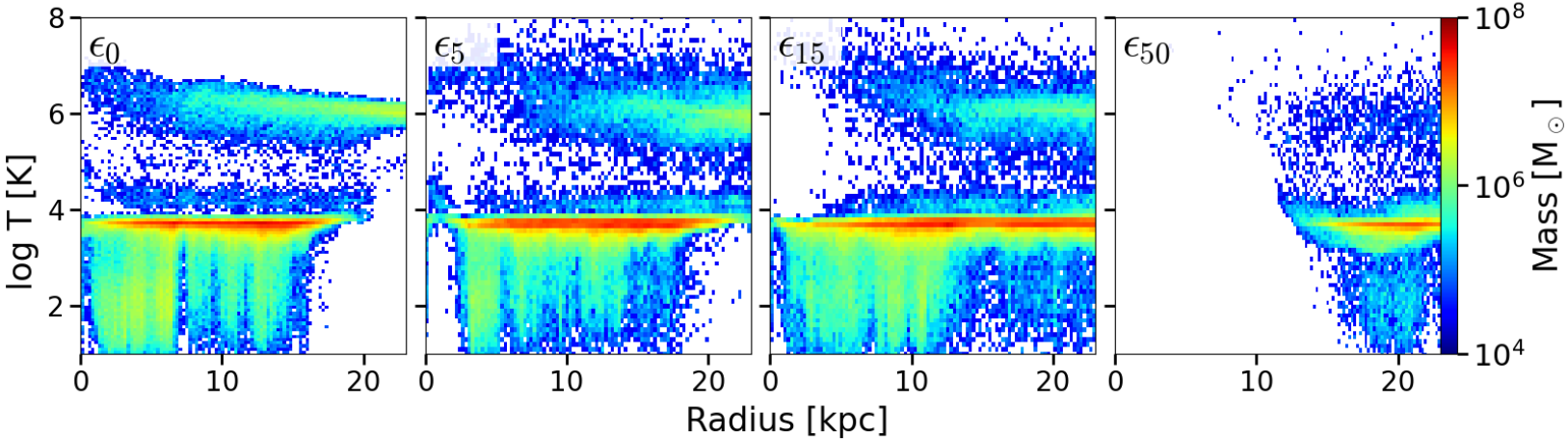}
\caption{Radial distribution of gas temperature within $0.1R_{\rm vir}$, at $z=0$. The temperature and radius are binned into $100\times 100$ cells. The color palette shows the mass within bins.  
\label{fig:TempMap}}
\end{figure*}

\begin{figure*}[ht!]
\includegraphics[width=1\linewidth]{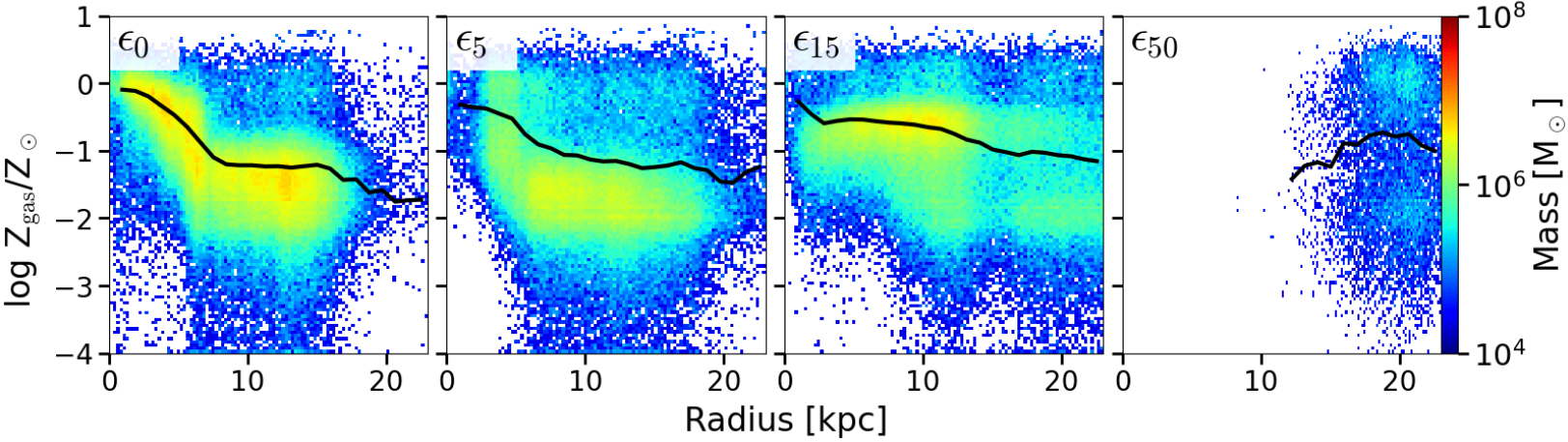}
\caption{As Figure\,\ref{fig:TempMap} but for the ISM metallicity. Black line is the mass-weighted average calculated in 1\,kpc wide spherical shells. The color palette shows the mass within the histogram bins. 
\label{fig:MetalMap}}
\end{figure*}

\subsection{The SMBH growth}
\label{sec:SMBH}

Table\,\ref{tab:BHprops} provides the final SMBH masses and other accretion parameters, including the jet mechanical luminosity, $L_{\rm jet}$, and the jet angle with respect to the inner 3\,kpc stellar disk spin axis. The final SMBH masses range within $M_\bullet\sim 5.4-7.2\times 10^7\,M_\odot$, so they differ by about 30\% from each other, increasing with $\epsilon$. The least efficient AGN model has a smaller final SMBH mass, thus we do observe that accretion rate correlates with the accretion efficiency $\epsilon$, as is explored further in Paper\,II. The simplest explanation for this effect is that the jet feedback is anisotropic in our models. Mostly, the jet points away from the galactic disk, so the SMBH can continue to accrete material even during periods of powerful outflows. The history of the accretion rate and growth by the SMBH, $\dot M_\bullet$, differ as well between the models. The accretion rate, averaged over the last 30\,Myr, declines sharply and monotonically from $\dot M_\bullet \sim 6\times 10^{-3}\,M_\odot\,{\rm yr^{-1}}$ for $\epsilon_5$ down to $\sim 5\times 10^{-4}\,M_\odot\,{\rm yr^{-1}}$ for $\epsilon_{50}$, i.e., by more than an order of magnitude.

The SMBH jet angle with the disk evolves by inheriting the angular momentum of the accreted gas.  The angle has been averaged over the last 30\,Myr for each model. The jet angle confirms that it points away from the disk, as the angle ranges between $21^\circ-52^\circ$.  The SMBH accretion rate at $z=0$, and thus the jet mechanical luminosity, appear to scale inversely with the accretion efficiency. So that the $\epsilon_{50}$ model has $L_{\rm jet}$ about an order of magnitude lower than the other two models, and in accordance with $\dot M_\bullet$. This result emphasizes that the current feedback by the SMBH does not reflect the long time history of the gas supply in the galaxy centers, as will be shown in Paper\,II.

\subsection{AGN feedback and the ISM}
\label{sec:ISM}

Unless otherwise specified, the ISM discussed here refers to all of the gas within 0.1$R_{\rm vir}$ for each of the models. As already seen in the final SFRs of Table\,\ref{tab:galprops}, the addition of the AGN feedback decreases the SFR in a galaxy with increasing $\epsilon$ at $z=0$, resulting in a factor of $\sim 3$ difference in the total stellar mass. 

The explanation to this result lies in the history of the SF in the modeled galaxies (Figure\,\ref{fig:agevsrad}).  While $\epsilon_0$ and $\epsilon_5$ models exhibit a gradual decline in the SFR over the cosmological times, the high $\epsilon$ models display dramatic declines after the SMBH has been seeded. The $\epsilon_{15}$ model nearly recovers its SFR after a couple of Gyrs, the $\epsilon_{50}$ model never recovers fully. It takes about 7\,Gyr to attain $\sim 10^{-3}\,M_\odot\,{\rm yr^{-1}}$ after $z\sim 0.3$. The dominant effect of the AGN feedback has been to reduce the SFR in a galaxy. 

While Table\,\ref{tab:galprops} shows that the total stellar mass of these galaxies differs by factor of $\sim 3$, their SFRs differ by orders of magnitude as a function of the efficiency. The SF history can be inferred from Figure\,\ref{fig:agevsrad} and differs dramatically over extended time periods. Why does such a difference in the SFR result only in a factor 3 difference in $M_*$? The explanation lies in Figure\,\ref{fig:agevsrad}.  As we have estimated in section\,\ref{sec:morph}, most of the stars in modeled galaxies have been formed before $z=1$. Moreover, half of these stars have been formed before the SMBH has been seeded in the galaxies, and hence did not experience the AGN feedback. 

Clearly, one should conclude on this point that the SMBH seeds should form in the early stages of the host galaxy evolution in order to have a more profound effect on the stellar population. This issue is further addressed in Paper\,II.   

Table\,\ref{tab:galprops} shows the final gas fraction in each of the models. At $z=0$, all four galaxies show a gas fraction of around $7-17\%$. In absolute numbers, some small dispersion exists in the gas masses. This dispersion exists also for stellar masses in galaxies. In relative numbers, the gas fraction in $\epsilon_{15}$ stands out, because it is normalized by a smaller stellar mass in this model.

Next, we analyze the gas temperature and metallicity in modeled galaxies. Figure\,\ref{fig:TempMap} displays the distribution of the gas temperature as a function of distance from the galaxy within 25\,kpc, i.e., $\sim 0.1R_{\rm vir}$, coloring it with mass.  All the models display a bimodal distribution of gas with temperature. The cold, $< 10^4$\,K starforming gas is present in all galaxies, but its radial distribution differs substantially. In the $\epsilon_0$ model, it extends from the very center to the outskirts of the galaxy, $\sim 20$\,kpc, with most of the gas residing at the cooling floor of atomic hydrogen (although molecular and metal cooling are active). Already for $\epsilon_5$ and $\epsilon_{15}$, the cold gas does not extend to the center --- the amount of $\sim 10^4$\,K gas decreases at small radii. For the $\epsilon_{50}$ model, the cold gas is essentially pushed outside 15\,kpc. 

\begin{figure}[ht!]
\centering
\includegraphics[width=0.85\linewidth]{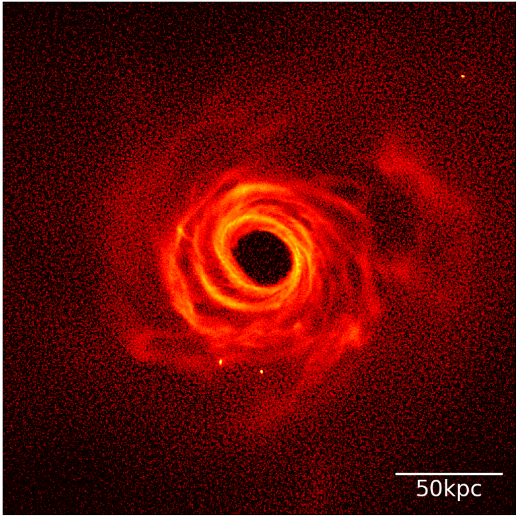}
\caption{Projected view of the face-on gas disk in $\epsilon_{50}$ galaxy at $z=0$. Note the large cavity introduced by the AGN feedback. The gas is pushed out of the HOP galaxy, so only a thin ring remains attached, as shown in Figures\,\ref{fig:galaxies} and \ref{fig:gasDisk}. The gas outside this ring belongs to the CGM.   
\label{fig:e50_large_scale_gas}}
\end{figure}

\begin{figure*}[ht!]
\center
\includegraphics[width=0.8\linewidth]{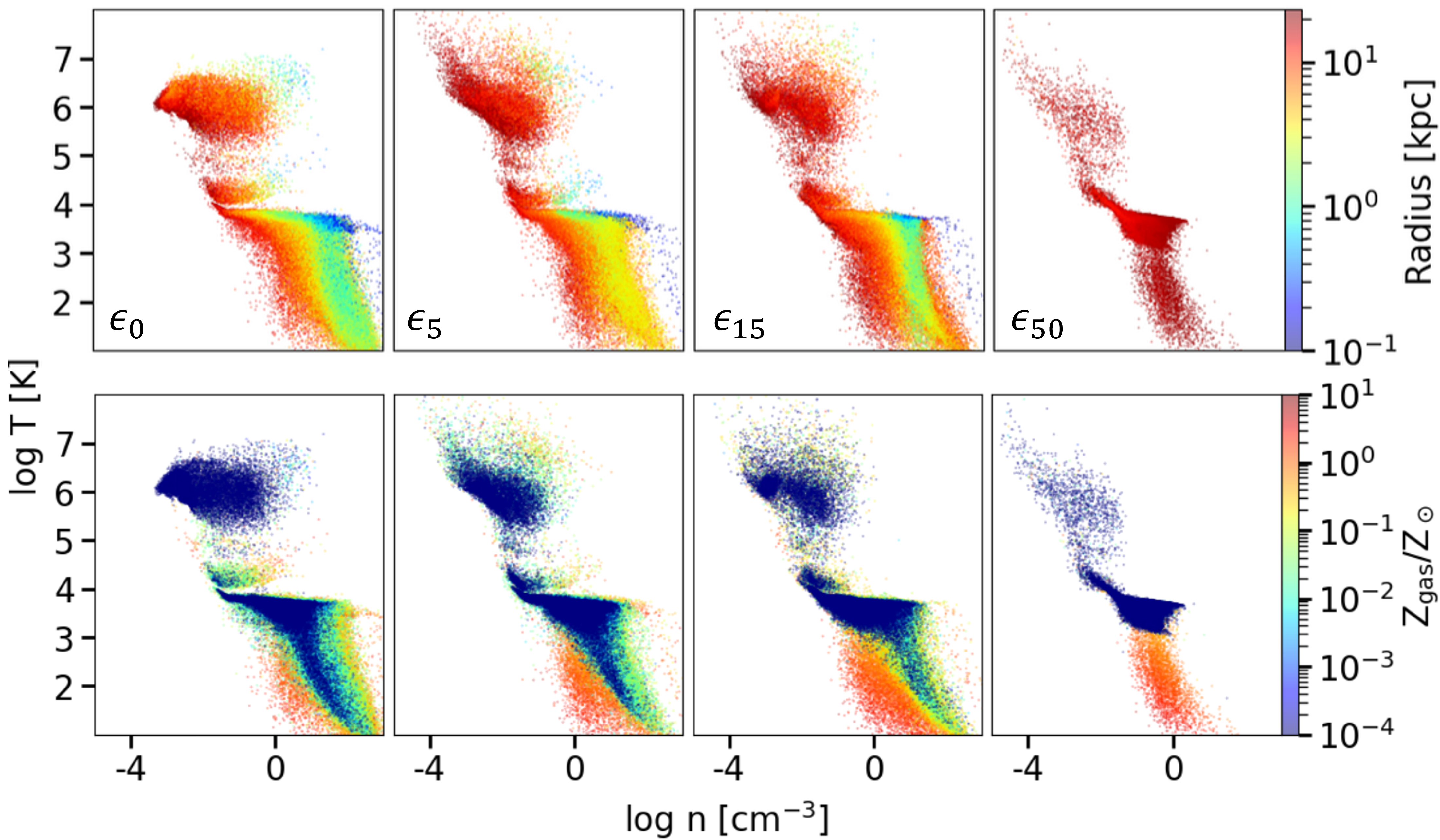}
\caption{Temperature-density phase diagrams for modeled galaxies within $0.1R_{\rm vir}$ at $z=0$. The top row is colored by the particle distance to the galaxy center. The bottom row shows the same ISM colored by its metallicity. 
\label{fig:ISMrhoT}}
\end{figure*}

The hot gas with $T\gtorder 10^6$\,K extends to the central regions in the $\epsilon_0$ model.   Not much gas resides in the intermediate (warm) temperature between the cold and hot gas. For $\epsilon_5$ model, this trend continues. The hot gas has a larger distribution in $T$, which extends even beyond $10^7$\,K. For stronger accretion efficiency, the hot gas is pushed further out in radius, beyond 10\,kpc, its temperature has decreased below $10^7$\,K. Its mass within the HOP galaxy is very small, as a result. This gas extends well outside the HOP galaxy. 
 
\begin{figure*}[ht!]
\center
\includegraphics[width=0.8\linewidth]{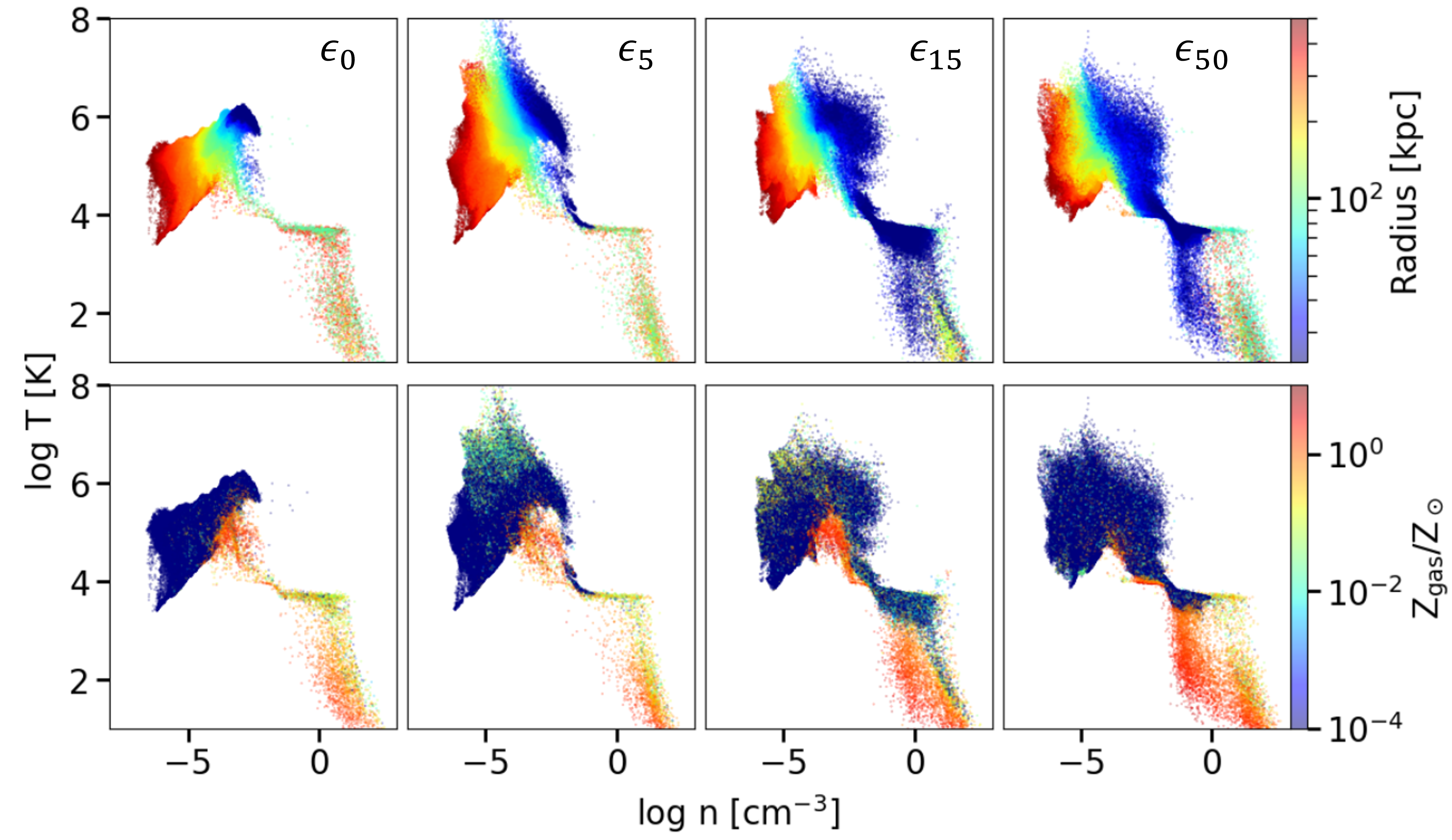}
\caption{As Figure\,\ref{fig:ISMrhoT} but showing all the gas in the range of $0.1R_{\rm vir}-2R_{\rm vir}$, representing the CGM only. Note, however, that cold gas residing in the substructure was not removed --- it is seen in all models as a distant high metallicity gas.
\label{fig:CGMrhoT}}
\end{figure*}

We turn to the ISM metallicity (Figure\,\ref{fig:MetalMap}). In all models, the dominant gas mass has metallicity of $Z/Z_\odot\sim 10^{-3}-1$. But the radial distribution and the mass of the gas which has this metallicity differs between the models. The higher metallicity in the central regions is an indication of an inside-out enrichment. The majority of the gas in $\epsilon_0$  and $\epsilon_5$ models, displays the radial gradient of metallicity in the central few kpc. Beyond 10\,kpc, most of the gas has $Z/Z_\odot\sim 3\times 10^{-3}-10^{-1}$. 

The $\epsilon_{15}$ model is well mixed radially, with most of the gas having $Z/Z_\odot\sim 3\times 10^{-2}-0.3$ inside 10\,kpc. Beyond this radius, the distribution is bimodal, with $Z/Z_\odot\sim 10^{-2}$ and $10^{-3}$. Although pushed out, the gas in the $\epsilon_{50}$ model displays the same bimodality around $Z/Z_\odot\sim 1$ and $10^{-2}$. 

Note that we do not explicitly model metal diffusion, and metals are distributed to the gas through SN feedback only. So, is the bimodality an artifact? To follow up, we have analyzed the gas distribution in $\epsilon_{15}$ and $\epsilon_{50}$ models and determined that all models have an offset outer gaseous ring of a low-metallicity accreting material. Recent star formation in this pristine material accreting from the filaments appears to contribute to the gradient and bimodality observed at large radii for these models.

To emphasize the presence of an extended gaseous disk in $\epsilon_{50}$ galaxy, we show the gas distribution within a 240$\times$ 240\,kpc region in Figure\,\ref{fig:e50_large_scale_gas}. This gas does not appear in the HOP-selected galaxies shown in Figures\,\ref{fig:galaxies} and \ref{fig:gasDisk}, with the exception of a narrow ring surrounding these galaxies. This extended disk in tandem with the central cavity displays the effect of the AGN feedback. While the stellar component in such galaxies is still present and displays a normal disk, the SF has been quenched there, and their colors would represent those of the lenticular galaxies (This point will be addressed in detail in Paper\,II). 

In order to investigate these effects further, we have plotted the temperature-density phase diagram for the ISM by coloring it in two-ways (Figure\,\ref{fig:ISMrhoT}). The top row shows the phase diagram with colors representing the radial distance from the galaxy center \citep[e.g.,][]{bi24}. The bottom row displays the same gas colored by its metallicity.  These frames exhibit a progressively reduced higher density gas, especially at $\gtorder 10$\,kpc, while lower frames display lower metallicity in the higher density gas.  

The top frame of Figure \ref{fig:ISMrhoT} shows that a lower temperature of the ISM is found at higher density, and that the higher temperature is found at larger radii. The bottom frame in this figure offers another view of the gas metallicity, as it relates the temperature and density distributions in each galaxy colored by metallicity. We see that the amount of the highest metallicity gas decreases with $\epsilon$ and tends to be `pushed' to lower densities. We also observe a decrease in low-metallicity gas at high densities (and lower temperature), and a decrease of low-metallicity gas at high temperature, which becomes more diffuse with increasing accretion efficiency.

\begin{figure*}[ht!]
\includegraphics[width=1\linewidth]{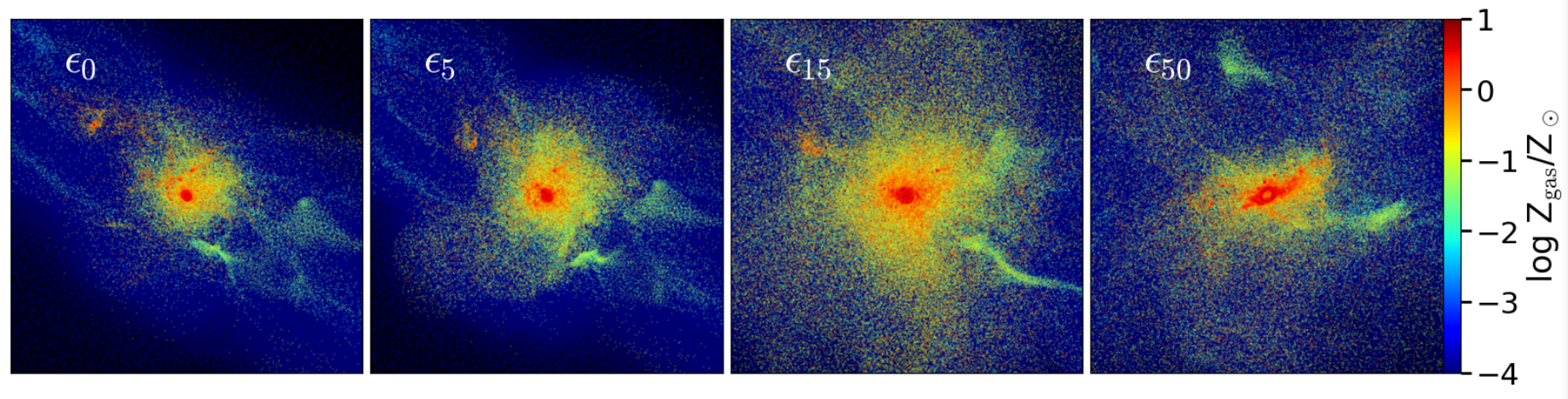}
\includegraphics[width=1\linewidth]{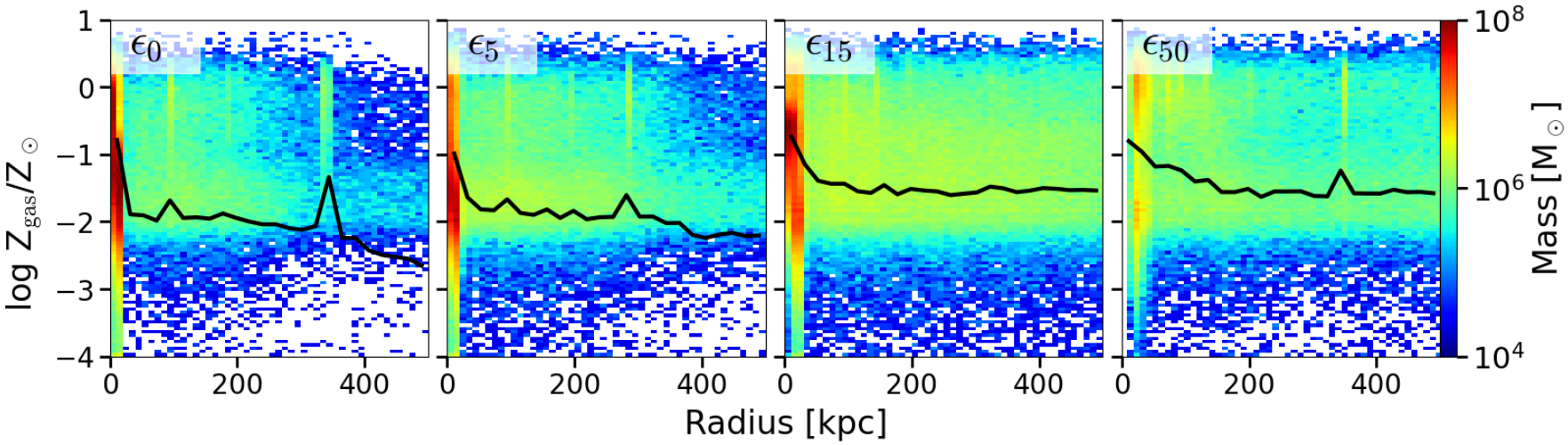}
\caption{Gas metallicity in the CGM within $2R_{\rm vir}$. {\it Top row:} shows the gaseous component of the CGM with each gas particle colored by its metallicity. {\it Bottom row:}  2D histogram of gas metallicity vs radius. The mass-weighted average metallicity calculated in 20\,kpc wide spherical shells is over-plotted.  The color palette shows the mass in each pixel in this phase space.  
\label{fig:MetalMapCGM}}
\end{figure*}
\newpage

\subsection{AGN feedback and the CGM}
\label{sec:IGM}

We turn now to the gas properties outside the galaxies, analyzing a good part of the CGM. The top row of Figure\,\ref{fig:CGMrhoT} displays the CGM gas diagram within the range of of $0.1R_{\rm vir}- 2R_{\rm vir}$, which is colored similarly to Figure\,\ref{fig:ISMrhoT}, i.e., by the distance of the individual gas particles to the center of the main galaxy.  Note that the gas residing in substructures was not removed and can be seen in $\epsilon_{15}$ and $\epsilon_{50}$ models at large distance from the central galaxy as a high metallicity gas.

We can see that the presence of an AGN feedback increases the amount of hot, $T > 10^4$\,K, low-density gas across the shown CGM, and that the density of this gas decreases monotonically with increasing distance. We also see that there is more cold, $\ltorder 10^4$\,K, gas just outside of $0.1R_{\rm vir}$. 
 
In the bottom row, the gas is colored by metallicity.  This distribution displays a progressively larger density range of the gas with $T\ltorder 10^4$\,K with accretion efficiency.  The population of the cold gas near the galaxy, $\sim 0.1R_{\rm vir}$, is split into two parts.  The portion of the gas that sits near $0.1R_{\rm vir}$, is $\sim 10^4$\,K, and has a very low metallicity must be either recently accreted or actively accreting from the cosmological filaments. There is also $\ltorder 10^4$\,K gas that seems to be quite high in metallicity, with $Z_{\rm gas}/Z_{\odot} > 1$, and sits just outside the starforming disk, outside the HOP galaxy. This high metallicity gas is present in both the $\epsilon_{15}$ and $\epsilon_{50}$  models, and serves as further evidence for the extension of the gaseous component of the galaxy being pushed out due to the presence of a stronger AGN feedback. Note that alongside this population is the cold gas at large distances, this sits inside the substructure, which does not belong to the CGM technically. Also, we observe that the hot, low-density gas displays an intermediate to high metallicity, and so was likely ejected from the galaxy by the AGN feedback. Therefore, the presence of an AGN serves to both heat and enrich the CGM, and to extend the reach of the gaseous component in a galaxy. 

Figure\,\ref{fig:MetalMapCGM} provides the CGM metallicity distribution within $2R_{\rm vir}$. The top row displays snapshots which emphasize the increase in reach of the high metallicity region with $\epsilon$. At the same time, the $\epsilon_{50}$ model reveals the change in shape of this region. It becomes elongated horizontally, approximately along the jet orientation. 

The bottom row displays the radial distribution of metallicity, with the mass-weighted average superposed. It confirms an increase in the mass-weighted average metallicity with AGN feedback. It also shows that the central metallicity gradient becomes shallower with $\epsilon$, a sign of increased radial mixing.

We summarize the halo baryonic content in Table\,\ref{tab:cgm}, which contains a fraction of the CGM. The stellar fraction shows a decline with increasing efficiency, almost by a factor of 2. The gas fraction remains steady, except for the $\epsilon_{50}$ model which drops significantly. Overall, the baryonic contents decrease with $\epsilon$ --- a clear signature of the AGN feedback.

\begin{deluxetable}{cccccccc}[ht!]
\tabletypesize{}
\tablecolumns{5}
\tablecaption{Halo Baryonic Content at $z=0$} \label{tab:cgm}
\tablehead{
\colhead{Model} & \colhead{log\,$M_{\rm halo
}/M_\odot$} & \colhead{$f_{\rm h,*}$} & \colhead{$f_{\rm h,gas}$} & \colhead{$f_{\rm h,baryons}$}  \\
\colhead{$\epsilon$} &  &  & &  
}
\startdata {$\epsilon_0$} & 11.92 & 0.13 & 0.04 & 0.17 \\
{$\epsilon_5$} & 11.91 & 0.13 & 0.04 & 0.17 \\
{$\epsilon_{15}$} & 11.88 & 0.08 & 0.04  & 0.12 \\
{$\epsilon_{50}$} & 11.87 & 0.07 & 0.03  & 0.10 \\
\enddata
\tablecomments{All quantities have been calculated within the virial radius of the DM halo.  In this case, $M_{\rm halo}$ is the total DM$+$baryonic mass within $R_{\rm vir}$. {\it Columns:} (1) Model name, (2) $M_{\rm halo}$ mass, (3) $f_{\rm h,*}$ fraction of stellar mass, (4) $f_{\rm h,gas}$ fraction of the gas mass, (5) $f_{\rm h,baryons}$ fraction of baryons.}
\end{deluxetable}

\subsection{The jet cocoons and the ISM, CGM and IGM}
\label{sec:CGM}

The jet cocoons in our modeled galaxies propagate well beyond the HOP-galaxies, and, depending on the AGN feedback, even cover the region beyond the host halo virial radius. We follow the definition  of the circumgalactic medium (CGM) as contained within $4R_{\rm vir}$. This distance corresponds roughly to the baryonic backsplash radius, where some of the inflow misses the central galaxy and is diverted out entirely because of gravitational effects \citep{bi24}, see also \citet[][and refs. therein]{tumlinson17}. The complicated kinematics of the CGM, which includes the filamentary and diffuse inflow, as well as outflow generated by the backsplash, and, in our case, also by the AGN-driven outflow, provides a challenge to disentangle all these phenomena. Beyond the CGM, the AGN outflow propagates throughout the IGM.  

The AGN jet can deposit its linear momentum  {as well as thermal} and kinetic energy when interacting with the ambient gas. Our numerical simulations do not resolve the jet's head (the hot spot), where the jet energy is decollimated. However, we do resolve the backflow of the jet shocked particles and their interactions with the ambient gas --- these result in the formation of an overpressured and expanding cocoon. In order to investigate the interaction between this cocoon and the ambient gas within our modeled Seyfert galaxies and consequently its effect on the SF, we looked at the dynamic and thermodynamic properties of these cocoons. 

\begin{figure}[ht!]
\center
\includegraphics[width=0.8\linewidth]{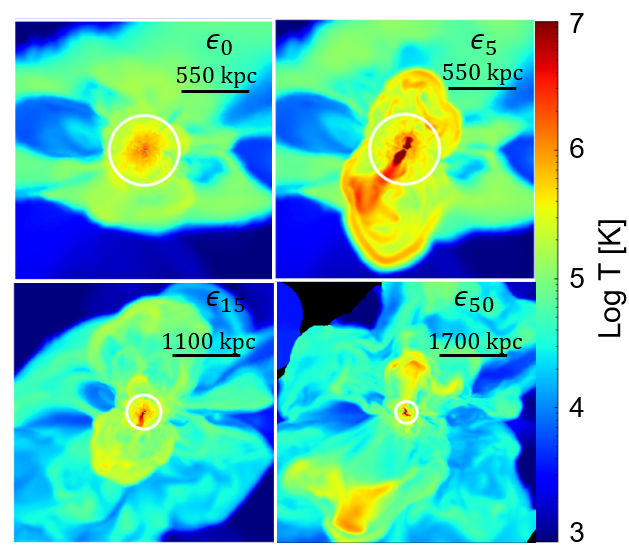}
\center
\includegraphics[width=0.9\linewidth]{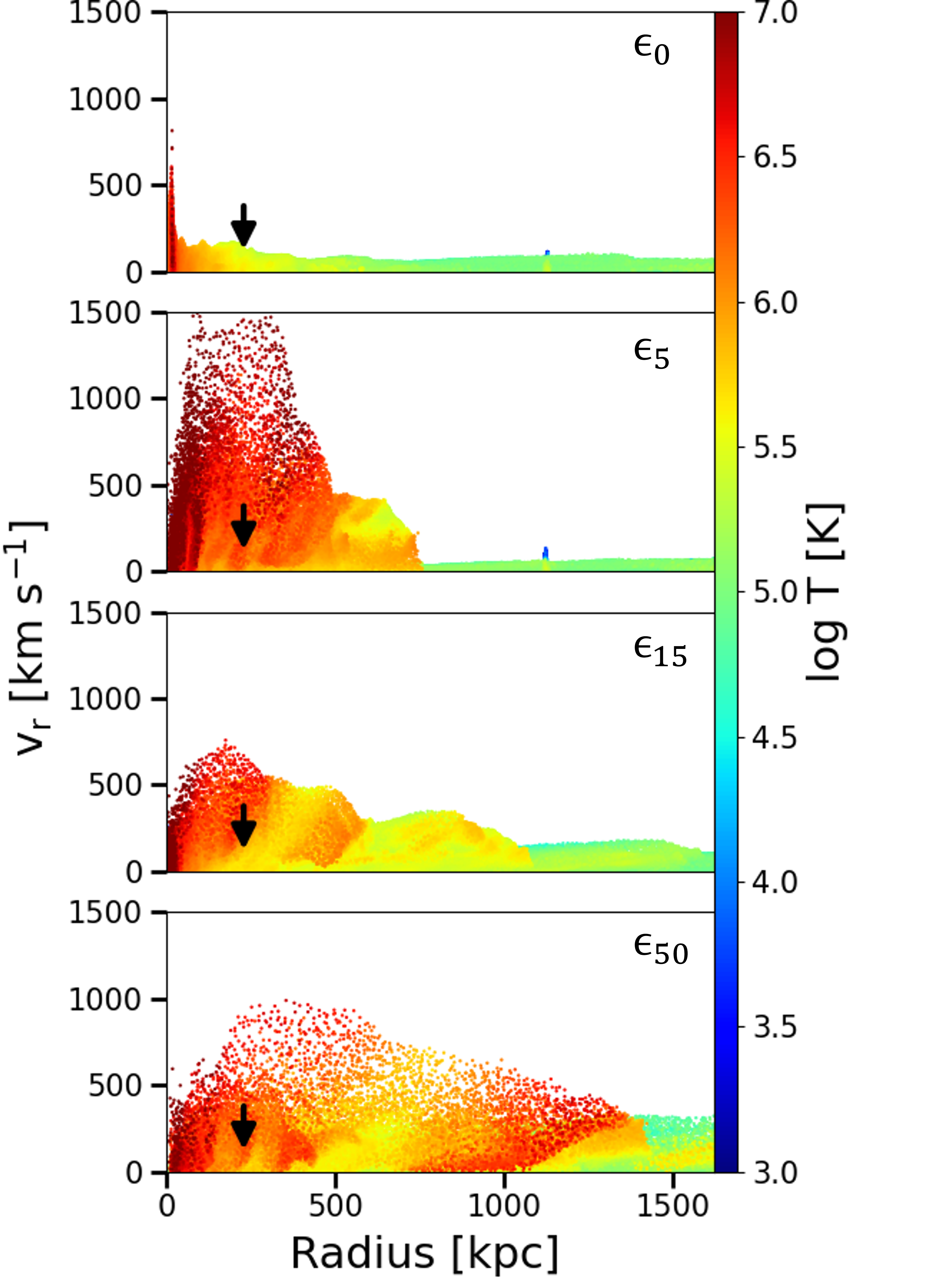}
\caption{Radial velocities and temperature of expanding cocoons in AGN and $\epsilon_0$ galaxies at $z=0$. {\it Top:}  slice projections of expanding cocoons into the CGM and IGM, with boundaries delineated by shocks. Each model is shown on a different scale, indicated by the scale bars, to emphasize the difference in the cocoon sizes. The color palette provides the temperature. The white circles are $R_{\rm vir}$. {\it Bottom:}  Jet cocoon expansion provided by radial velocity vs distance diagrams for the modeled galaxies. The black arrows mark the respective $R_{\rm vir}$. The color palette provides the gas temperature. Large regions of hot, fast outflowing velocity gas show radial propagation of the jets and jet cocoons. For $\epsilon_{15}$ and  $\epsilon_{50}$ models, previous generations of expanding cocoons have moved beyond the plotted region, affecting the CGM and IGM (not shown here).
\label{fig:cocoon}}
\end{figure}

\begin{figure}[ht!]
\includegraphics[width=1\linewidth]{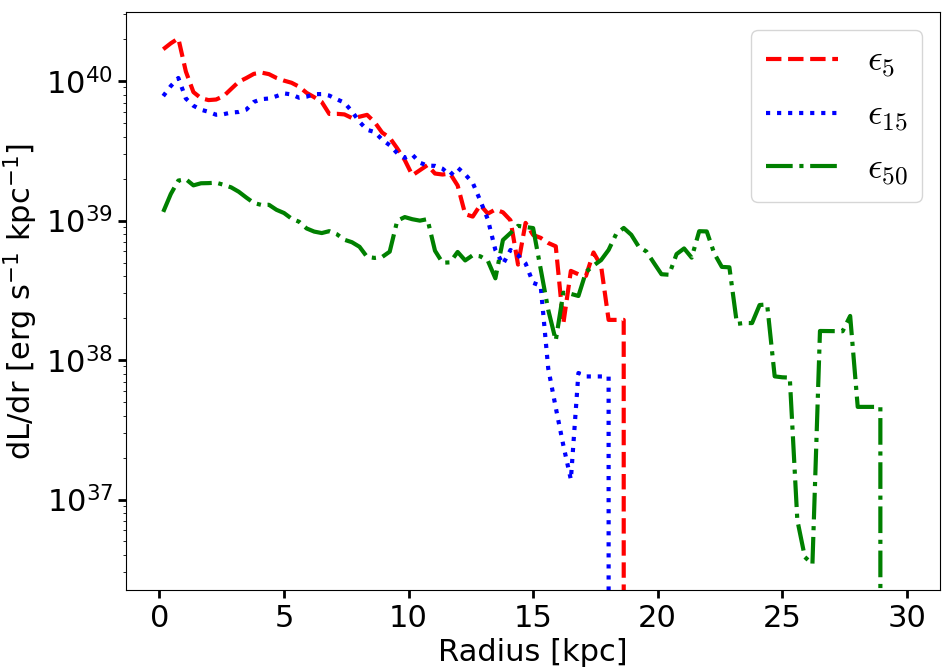}
\caption{ Total (thermal$+$kinetic) energy deposition rate into the environment by the jet particles averaged over the last $\sim 1$\,Gyr preceding $z=0$. The radius is binned in spherical shells with the bin size of 0.3\,kpc. Each jet particle has been followed from the point of emission until it's energy decreases to 10\% of its spawned value  or it merges with the surrounding gas, for each model with the SMBH. 
\label{fig:jetE}}
\end{figure}

In the $\epsilon_0$ model, only the SN feedback exists and affects the ISM and CGM within the virial radius by increasing mixing and temperature, as seen in the top frames of Figure\,\ref{fig:cocoon}. The ISM gas is heated, but is able to cool down, and only a small fraction has $T\sim 10^6-10^7$\,K. By $z=0$, the majority of the ISM has $T\ltorder 10^5$\,K. The effect of even weak AGN accretion efficiency, in the $\epsilon_5$ model, is more complex. The majority of the ISM still remains at $t\ltorder 10^5$, but the hot component is heated up to $T\sim 10^6-{\rm few}\times 10^7$\,K. The hottest gas within the CGM has $T\sim 10^5-10^7$\,K, and has obviously been affected by the interaction with the jet flow. Within the cocoon volume, most of the CGM has been heated up by a factor of $\sim 2-3$ compared to the $\epsilon_0$ model. 

The cocoon in $\epsilon_5$ model has stagnated at $\sim 750$\,kpc, so to $\sim 3R_{\rm vir}$. But the breakout has occurred only very late, at $z\sim 0.2$. The average velocity of the cocoon expansion is $\sim 260\,{\rm km\,s^{-1}}$, and the cocoon extends perpendicularly to the major filament where the galaxy resides, after the breakout. The velocity is clearly supersonic with respect to the ambient gas, and hence the cocoon expansion is delineated by a shock front, which are observable in the top frames of Figure\,\ref{fig:cocoon}. The cocoons are pushed by internal pressure which is supported by $T\sim 10^6-10^7$\,K.

In the $\epsilon_{15}$ and $\epsilon_{50}$ models, we detect multiple nested cocoons --- meaning that the outflow follows the same channel established by the preceding cocoon. In some cases the subsequent cocoon follows a different path, but after the breakout it always expands nearly perpendicularly to the main filament. In the $\epsilon_{15}$  model, the cocoon has reached 2\,Mpc at $z=0$ in its longest elongation. It has decelerated and stagnated at this distance. The average velocity of this cocoon is about $500\,{\rm km\,s^{-1}}$. 

The cocoon in the most energetic AGN $\epsilon_{50}$ model has reached $\sim 2.4$\,Mpc at $z\sim 0.8$, where it stagnated. Its initial speed exceeded $1000\,{\rm km\,s^{-1}}$, followed by the next generation cocoon which changed its direction due to a change in the jet orientation. The gas temperature behind the front of the cocoon is similar in all models. The timeline of the breakout, associated shock front velocities, and temperatures for the cocoons are quoted here for reference only. We reserve the details of the cocoon evolution for Paper II.

The lower frames of Figure \ref{fig:cocoon} show the radial velocities and temperatures of the gas particles in the cocoons at $z=0$.  It is evident that all AGN models feature multiple bursts of hot, fast moving, expanding gas on a scale of 1.5\,Mpc that are absent from the $\epsilon_0$ model.  The shock fronts delineate hot, $\sim 10^6-10^7$\,K, cocoon fronts, i.e., the shocks, that propagate away from the galaxy. The older cocoons, while hotter than ambient gas, have basically stopped their expansion. By comparing the radial occurrences of the hot outflows with the bubbles, we can see that they represent the same phenomena, which describe the jet cocoons.  

The highest velocities occur during the initial expansion of the cocoon ‘bubbles.’  Figure\ref{fig:cocoon} provides a snapshot at z=0 where only the $\epsilon_5$ model is undergoing this early breakout phase. These outflows have an effect on the CGM and IGM, enrich these regions with metals, and are expected to influence the accretion rate onto the central galaxy. Moreover,expanding cocoons should affect the ISM properties and the SF in the substructure surrounding the parent DM halos.   

While the cocoons reach a large distance from the galaxy, the majority of the hyper-refined jet particles typically propagate out to much smaller distances, subject to hydrodynamical interactions with the surrounding gas. The initial expansion of the cocoons is clearly driven by their over-pressure. Ram pressure can drive fast shocks which compress and sweep up the ambient gas. At large distances, the cocoon fronts appear to move inertially.  

The  mechanical energy output by the jets at $z=0$ is typical to that observed in Seyfert jets \citep[e.g.,][]{whittle04}, namely, $L_{\rm jet}\sim 2\times 10^{41}\,{\rm erg\,s^{-1}}$ in $\epsilon_5$,  $\sim 8\times 10^{40}\,{\rm erg\,s^{-1}}$ in $\epsilon_{15}$, and  $\sim 2\times 10^{40}\,{\rm erg\,s^{-1}}$ in $\epsilon_{50}$. This is about $\sim 10^{-5-6}$ of the Eddington mechanical luminosity of the modeled SMBHs.
 
We have analyzed the distribution of the deposited energy by the jet particles as a function of distance from the SMBH, shown in Figure\,\ref{fig:jetE}. It shows the  total (thermal and kinetic) energy  deposition rate into the ambient gas by the jet as a function of $R$  averaged over the last 1\,Gyr.  The energy deposition rate is found from the  total change in energy for each jet particle as it propagates away from the SMBH until it re-merges with the gas.  The changes in energy for all jet particles spawned within the 1 Gyr time period are summed and normalized by the radial bin size and time.  We find that nearly all of the thermal energy from the jet is injected when the jet particles merge with the surrounding gas. The jet particles barely escape beyond $\sim 0.1R_{\rm vir}$, and only in $\epsilon_{50}$ model.  The $\epsilon_{50}$ model shows that the deposited energy rate per unit radius is relatively flat out to $\sim$22 kpc, apart from a slight decrease in the inner few kpc. For the smaller efficiency models, dL/dR is a little steeper.

Note that most of the energy has been deposited in the central kpc at low $z$ by the $\epsilon_5$ and $\epsilon_{15}$ jets, while not by the $\epsilon_{50}$ jet, which deposits its energy more uniformly with $R$. This difference is apparently related to the gas distribution in modeled galaxies, where $\epsilon_{50}$ galaxy has a large central cavity devoid of gas at this time. The deceleration of the jet close to the production region has been suggested as the origin of the FR-I morphology in powerful radio sources by \citet{deyoung93}.

\section{Discussion and Conclusions}
\label{sec:discuss}

We used high-resolution zoom-in cosmological simulations to model evolution of three Seyfert-type galaxies, with the SMBHs seeded at $z=3.7$ and triggered collimated jets. We find that the SMBH jets in this energy range exert an effect on galactic morphology, their SFRs, their gas fraction, gas distribution, etc. The effects of this feedback, depending on the associated accretion efficiency, extend outside the galaxy, and influence the CGM properties even beyond $R_{\rm vir}$, up to the baryonic backsplash radii, determining the mass accretion onto the parent DM halo. The evolution of the AGN galaxies has been compared to a galaxy without the SMBH. We have analyzed the interactions between the SMBH jets, their host galaxies and immediate environment, the ISM and the CGM.  In this work we have only focused on the objects at $z=0$, while cosmological evolution for $z\ltorder 10$ will be presented in Paper\,II. 

 Feedback from jets launched by the SMBHs deposit both energy and momentum in the ambient gas. This process can have a potentially far reaching effect on the gas distribution in the host galaxy, its thermal and kinematic properties, and consequently on the SF. The integral effect on the evolution of Seyfert galaxies depends on a number of model parameters characterizing the jets. We choose to perform a controlled experiment by varying only a single parameter at a time. Here we present a suite of models which differ by Eddington fraction of the SMBH accretion rate. The SMBHs of $M_\bullet \simeq 10^6\,M_\odot$ have been seeded relatively late, when $M_{\rm vir}\sim 10^{11}\,M_\odot$.  The energy output by a seeded SMBH has been chosen as the thermal energy, kinetic energy, and linear momentum of jet particles emitted anisotropically along the spin axis of the SMBH.  Galaxies at $z=0$ differ in the following ways with increasing accretion efficiency, presented here along the $\epsilon$ sequence, from $\epsilon=0$ to $\epsilon=0.5$:

\begin{itemize}
\item Stellar masses in modeled galaxies within DM halos of $\sim 6.3\times 10^{11}\,M_\odot$ lie within the range of $3\times 10^{10}-10^{11}\,M_\odot$ --- the AGN galaxies have progressively smaller masses with increased $\epsilon$ compared to the $\epsilon_0$ galaxy. Both the stellar and gas fractions within the parent DM halos decrease along the $\epsilon$ sequence, from 12.9\% to 7.2\% (stellar) and 4.1\% to 2.8\% (gas), respectively. Hence, the baryonic fraction inside the virial radius decreases from 17\% down to 10\%.  

\item The radial distribution of the gaseous component  differs from model to model. Some of these differences can be understood following the cosmological evolution, others are a clear consequence of the AGN feedback. Increase in the size of the gaseous disk with increasing efficiency is the most obvious of such trends. The $\epsilon_{50}$ galaxy has nearly all the gas residing at the galaxy outskirts.   

\item Sersic decomposition of the stellar distribution shows the presence of a massive bulge in all galaxies, but $B/T$ decreases $\sim$1/2 along the $\epsilon$ sequence, from $\sim 0.55$ to $\sim 0.28$. The stellar disk mass and its scalelength decrease by a factor of $\sim 1.5$  and $\sim 2$, respectively, with $\epsilon$.

\item The SFR is inversely proportional to the accretion efficiency, which decreases the SFR by three orders of magnitude, from $2\,{\rm M_\odot\,yr^{-1}}$ to $2\times 10^{-3}\,{\rm M_\odot\,yr^{-1}}$. The $\epsilon_{50}$ model has essentially a quenched SF after $z\sim 1$. In all AGN models, stars forming after $z\sim 1$ have their origin outside the central few kpc. This central gap increases with $\epsilon$. The SFR in $\epsilon_0$ and $\epsilon_5$ models appear comparable, while $\epsilon_{15}$ is more similar to the $\epsilon_{50}$ model. Overall, the SF is shifted to larger radii with increasing $\epsilon$. 

\item The ISM metallicity is about 5 - 10 times smaller than the stellar metallicity, meaning that stars formed from a more metal-rich gas which has been subsequently replaced by the inflowing low-metallicity gas. The $\epsilon_0$ and $\epsilon_5$ models show a radial gradient in the average ISM metallicity, while $\epsilon_{15}$ and $\epsilon_{50}$ models are radially mixed.  

\item The CGM properties differ along the $\epsilon$ sequence. The cold, $\ltorder 10^4$\,K gas displays a progressively larger density range with $\epsilon$, while hotter gas shows a lower metallicity and a higher density range with increasing $\epsilon$. The mass-weighted average metallicity of the CGM gas is higher in $\epsilon_{15}$ and $\epsilon_{50}$ models by a factor of a few. 

\item All AGN models develop AGN-driven outflows, which define expanding cocoons (bubbles) outlined by shocks --- extending to $\sim 750$\,kpc in the $\epsilon_5$, to $\sim 2$\,Mpc in $\epsilon_{15}$, and to $\sim 2.4$\,Mpc in $\epsilon_{50}$ model. The cocoons are pressure-driven. The comparison $\epsilon_0$ model is subject only to the SN feedback which mixes the gas within $R_{\rm vir}$, and no cocoons have developed in this case. The cocoon expansion in AGN galaxies is supersonic with respect to the ambient gas, with $\sim 260\,{\rm km\,s^{-1}}$, $\sim 500\,{\rm km\,s^{-1}}$, and $\sim 1,000\,{\rm km\,s^{-1}}$, with increasing $\epsilon$. The radial velocities of the outflowing gas within the cocoons are much higher of course, reaching few$\times 10^3\,{\rm km\,s^{-1}}$. 

\item The cocoons observed in AGN galaxies have oval shapes, with axial ratios of 1:2 -- 1:3. The internal temperatures of the cocoons lie in the range of few$\times 10^5 - {\rm few}\times 10^7$\,K, where the upper limit delineates the front shocks in the expanding cocoons.

\end{itemize}
 
\subsection{Comparison with observations}
\label{sec:compare}

We now turn to comparison with observations and numerical modeling, and quote references in addition to those discussed in Section\,\ref{sec:intro}. Observational data on jets and their feedback in Seyfert galaxies is scarce, but interest is growing. We are still lacking multi-wavelength, high-resolution statistical surveys which can address the effects that these pc- and kpc-scale jets have on their host galaxies, which are disk-type, unlike their more powerful radio galaxies counterparts. Currently, the comparison is mostly limited to individual objects  \citep[e.g.,][]{Varglund2022}.

Observations of jetted Seyfert galaxies show that they typically reside in disk galaxies with spiral morphologies and host SMBHs in the mass range of $10^6-10^8\,M_\odot$ \citep[e.g.,][]{vietri24} --- our modeled galaxies are bulged disks and their SMBHs fall within this range (see Table\,\ref{tab:BHprops}).
 
There is abundant evidence that Seyfert jets generate bow shocks, through interactions with the ISM, that propagate out with velocities up to $10^3\,{\rm km\,s^{-1}}$ \citep[e.g.,][]{cecil00,Tadhunter2014,Murthy2022,peralta23,Duggal2024}. Multiple generations of misaligned jet radio lobes have been observed, indicating multiple episodes of jet activity \citep{sebastian2019,rao23}.

Studies of CO emissions or kinematics of atomic and molecular gas in jetted low-luminosity AGN have shown that the gas is often displaced from the nuclear region, forming cavities, or rings, that can reach radii of $\gtorder 6$\,kpc or clouds of displaced gas \citep{garcia-burillo24,Nesvadba2021,nandi23,matsushita07}. These rings have been shown to be associated with either decreased SF \citep{Nesvadba2021,rao23,nandi23} or localized regions of enhanced SF \citep{Duggal2024,venturi21,krause23}. Gaseous rings can be associated with jets interacting with the MHD winds associated with the BEL regions and molecular tori \citep[][]{blandford82,emmering92,konigl94,bottorff97,elitzur06} or with the gas elevated above the disk plane by SN feedback, and are not required to be directed into the disk plane. Also, the disk plane can be warped, as in our modeled galaxies. Our results confirm these observations as at z=0, but our jets do not impact the disks directly. We observe regions with damped SF around the SMBHs, formation of central cavities and extended gaseous and stellar disks with increasing accretion efficiency (e.g., Figs.\,\ref{fig:profile} --- \ref{fig:gasDisk}).  

Looking into the SF distribution, \citet{acharya24} show that disk-dominated and pseudo-bulge AGN-hosting galaxies tend to have a decrease in SF in the central region and enhanced SF at larger radii, leading to an overall SF matching that of similar star-forming galaxies.  This result is further supported by other observations \citep{Kurian2024,lammers23,krause23}.  Bulge dominated systems have a flat SF distribution across galactic radii.  The central gap which is mostly devoid of SF after $z\sim 1$ is observed in all of our AGN galaxies, and clearly correlates with the strength of the AGN accretion efficiency.  

A review by \citet{krause23} emphasizes how CGM density and virial temperature can strongly influence the ability of the jet to propagate outward and whether one finds compression and associated triggered SF, or heating and gas expulsion resulting in quenching of SF.

Increased radio emission strength, along the sequence from star-forming galaxies to composite (i.e., galaxies with radio emission from SF and jets) to Seyfert and LINERs, has been claimed in \citet[][]{vitale15}. This increase appears to correlate with an analogous decrease in H$\alpha$ emission, which hints that jets are indeed instrumental in halting SF not only in large elliptical galaxies, but also lower mass disk galaxies.  

\citet{varglund23} has studied colors and morphologies of a sample of NLS1 galaxies and found disky morphologies and increased dust extinction in the centers.  The latter feature is often seen in $\gamma$-ray-detected jetted NLS1 galaxies. This reddening has been correlated with the jet activity in other observations \citep[e.g.,][]{olguin-iglesias20,berton19}.

Finally, \citet{cunlow04} observed an increased radial mixing of metals in stellar populations of Seyfert galaxies in comparison to non-active galaxies, which show enhanced metallicity in stars at the center. Possibly also a reduced age gradient for Seyferts. 

\subsection{Comparison with simulations}
\label{sec:compare numerics}
Numerical work on the effects of AGN jet feedback in Seyfert-type galaxies over cosmological timescales is currently limited, perhaps in anticipation that AGN feedback has little effect on MW-mass and smaller galaxies. Building up a statistical sample of observational and theoretical works is essential to understanding the full impact of low-luminosity AGN jets on their hosts and for lying groundwork for high-$z$ observations of these galaxies.

The aim of our simulation set has been to evaluate the impact of varying efficiency of the SMBH accretion in Seyfert galaxies. Isolating the effect of this parameter in a low-luminosity AGN model along with collimated jet feedback has not been addressed so far in the literature for cosmological simulations. Numerical simulations have demonstrated the ability of AGN jets to halt the cooling flows in massive galaxies and galaxy clusters. Some works point that jets act to reduce the stellar mass in MW-type and smaller host galaxies as well \citep{wellons23,byrne24, irodotou22, okamoto08}. However, short-term studies have noted brief or localized increases in the SFR in MW-mass, or Seyfert-type, galaxies \citep{mukherjee18,feliz23}. 

We observe that SF is pushed to large radii, due to a more extended gaseous disk with an increased accretion efficiency.  \citet{irodotou22} observed the same trend but using a ‘radio’ mode feedback, by dumping thermal energy in the CGM. \citet{byrne24} shows that central gas surface density decreases and stellar half-mass radii tend to grow by adding additional feedback in the form of jets, radiation, and cosmic rays.

Formation of central gaseous voids, as observed in our highest efficiency model, $\epsilon_{50}$, is also reported by \citet{qutob23}. Short-term (non-cosmological) studies of jets, which vary their orientation with respect to the galaxy, show that the jet can displace gas and dissolve cold clouds in the central regions of galaxies \citep{mukherjee18, talbot22}.

Using the Sersic decomposition, we show that increasing SMBH accretion efficiency leads to a decrease in B/D, B/T, in disk and bulge masses and scale length,  giving an altogether smaller bulge component and a ‘fluffier’ disk.  Adding the `radio' mode does not lead to the same outcome \citep{irodotou22}. Kinematic decomposition finds that the strongest and combined feedback modes lead to dissolution of the disk component, but that intermediate feedback shows little change from the fiducial models \citep{byrne24, okamoto08}. While not quantitatively explored in the paper, \citet{wellons23} mentioned that various AGN models tested do lead to differences in the host morphology.

Exploration of long-term effects to metallicity and mixing in simulations of Seyfert galaxies is lacking.  While we do not model the metal diffusion explicitly, we observe flatter final radial metallicity profiles with increased accretion efficiency, as well as slightly increased metallicity in parent halos. This effect is also noticed by \citet{Appleby21}, they observed an increase in the metallicity of hot CGM gas in full-box SIMBA simulations which include jet feedback. \citet{talbot24} similarly find increased metallicity in the CGM due to the presence of jets, and so did \citet{qutob23}, but their jet feedback included cosmic rays and radiation.

Modeling the evolution of jet cocoons in Seyfert galaxies under the influence of variation in accretion efficiency, and specifically the associated timescales, shapes, shocks, internal velocities and temperature, has not been addressed in the literature so far. However, short-term studies about launching and propagation of cocoons from low-luminosity AGN do exist. \citet{talbot24} found comparable maximal velocities and temperatures inside the cocoon, but for higher luminosity jets in merging galaxies. \citet{qutob23} observed cocoon properties scaling with increasing a multi-component feeback.  \citet{tanner22} studied the effect of varying the inclination angle of a low-luminosity jet with respect to the galactic disk.

In summary, similar trends and effects which show up in our simulations of Seyfert galaxies at $z=0$ have been observed in nearby Seyfert galaxies and simulation studies. Although statistical analysis of these properties is currently absent, our results point to the need for a systematic study of feedback from low luminosity AGN, like Seyfert galaxies.  In our study, we have limited the parameter space to modifying only the efficiency of the SMBH accretion which has a profound effect on the ISM of the host galaxy, whose properties will determine the morphology of the underlying stellar disk. In the subsequent paper\,II, we analyze evolution of Seyfert-type galaxies for $z\ltorder 10$, and focus on effects of the SMBH seeding redshift. 
\newpage
 
\section{Acknowledgments}
We thank Phil Hopkins for providing us with the latest version of the code, and to Alessandro Lupi, Kung-Yi Su, and Paul Torrey for their help with GIZMO. Many thanks to Da Bi and Xingchen Li for invaluable help with many aspects of the work carried out here. I.S. is grateful for a generous support from the International Joint Research Promotion Program at Osaka University, and acknowledges the hospitality of KITP where part of this research has been conducted. The KITP is supported by NSF PHY-1748958.  E.R.D. acknowledges support of the Collaborative Research Center 956, subproject C4; and from the Collaborative Research Center 1601 (SFB 1601 sub-project C5), both funded by the DFG–500700252. This work used Expanse CPU at San Diego Supercomputer Center (SDSC) through allocation PHY230135 from the ACCESS program, which is supported by NSF grants 2138259, 2138286, 2138307, 2137603, and 2138296 \citep{boerner23}, and by the University of Kentucky Lipscomb Computing Cluster. We thank Vikram Gazula at the Center for Computational Studies (UK) for a very generous computational support throughout this project.
 
\section*{Data Availability}

The data used for this paper can be available upon reasonable request. 

\bibliographystyle{aasjournal}
\bibliography{PaperI}{}

\appendix
\section{Sersic decomposition}\label{sec:appI}
\label{sec:append}

 
The bulge and disk components of the galaxy have been determined by fitting the 1-D face-on stellar surface density profile of the stellar component of each galaxy within 0.1$R_{\rm vir}$, and limited to $\pm$5 kpc in z, to a model that combines the Sersic function for the bulge, and two exponential disks, namely, the inner and outer disks:

\begin{equation}
\Sigma(R)=\Sigma_{\rm e}e^{-b_{\rm n}[(R/R_{\rm e})^{1/n}-1]}+\Sigma_{01}e^{-R/R_{\rm disk1}}+\Sigma_{02}e^{-R/R_{\rm disk2}},
\end{equation}
where  $\Sigma_{\rm e}$ and $R_{\rm e}$ are the effective surface density and radius of the bulge, and $n$ is the bulge Sersic index. The best fit has been achieved using a double-exponential disk. The fitting parameters are the disk's central surface densities $\Sigma_{01, 02}$ ,and their scalelengths $R_{\rm disk1, disk2}$.

\begin{figure}[ht!]
\center
\includegraphics[width=0.5\linewidth]{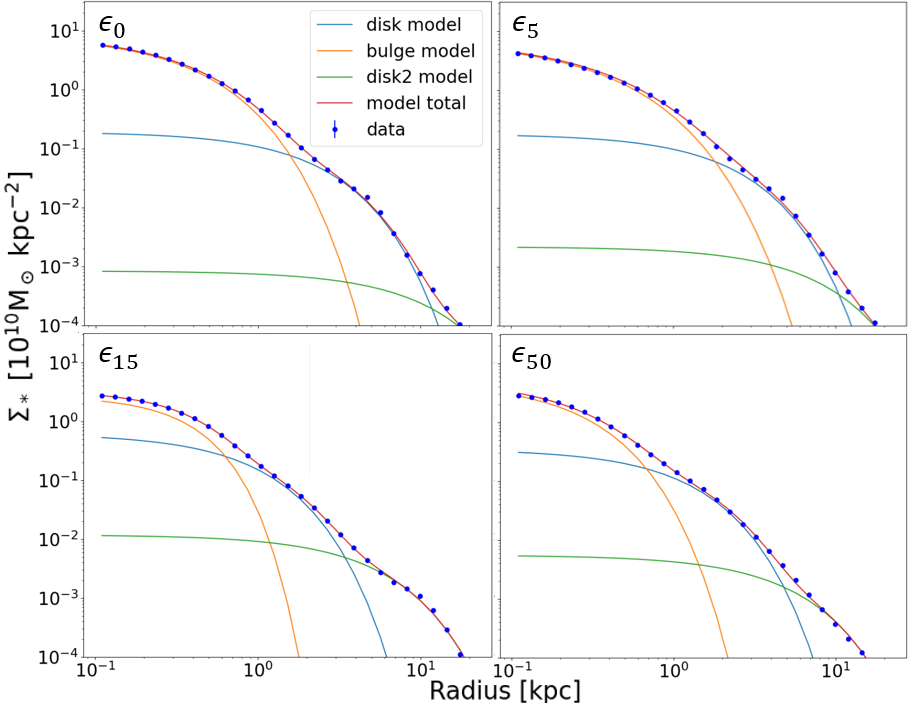}
\caption{The Sersic decomposition of the stellar component in modeled galaxies within $0.1R_{\rm vir}$. The total profile fits to the face-on stellar surface density are shown for each model.  The top left is the $\epsilon_0$ model, $\epsilon_5$ the top right, $\epsilon_{15}$ on the bottom left, and $\epsilon_{50}$ on the bottom right.  
\label{fig:ssd}}
\end{figure}

The numerical values of all the parameters of the decomposition are given in Table\,\ref{tab:deluxesplit}.

\end{document}